\documentclass[english]{IEEEtran}
\usepackage[T1]{fontenc}
\usepackage[latin9]{inputenc}
\usepackage{verbatim}
\usepackage{float}
\usepackage{amsthm}
\usepackage{amsmath}
\usepackage{amssymb}
\usepackage{graphicx}
\usepackage{multirow}
\usepackage{color}
\usepackage{url}
\usepackage{slashbox}

\makeatletter

\providecommand{\tabularnewline}{\\}
\floatstyle{ruled}
\newfloat{algorithm}{tbp}{loa}
\providecommand{\algorithmname}{Algorithm}
\floatname{algorithm}{\protect\algorithmname}

\numberwithin{equation}{section}
\numberwithin{figure}{section}
\theoremstyle{plain}
\newtheorem{thm}{\protect\theoremname}[section]
\theoremstyle{definition}
\newtheorem{defn}[thm]{\protect\definitionname}
\theoremstyle{remark}

\theoremstyle{plain}
\newtheorem{lem}[thm]{\protect\lemmaname}
\newtheorem*{lem*}{Lemma}
\newtheorem*{rem*}{Remark}
\theoremstyle{remark}

\theoremstyle{plain}
\newtheorem{corollary}[thm]{\protect\corollaryname}
\theoremstyle{plain}
\newtheorem{proposition}[thm]{\protect\propositionname}

\@ifundefined{showcaptionsetup}{}{%
 \PassOptionsToPackage{caption=false}{subfig}}
\usepackage{subfig}
\makeatother

\usepackage{babel}
\providecommand{\claimname}{Claim}
\providecommand{\definitionname}{Definition}
\providecommand{\lemmaname}{Lemma}
\providecommand{\remarkname}{Remark}
\providecommand{\theoremname}{Theorem}
\providecommand{\corollaryname}{Corollary}
\providecommand{\propositionname}{Proposition}
\newcommand{\diag}{\operatorname{diag}}

\definecolor{PUorange}{cmyk}{0, 0.6, 1, 0}
\newcommand{\STFT}{\operatorname{STFT}}

\newcommand{\STFTdagger}{\operatorname{STFT}^{\dagger}}

\begin{document}

\title{Non-Convex Phase Retrieval from STFT Measurements}

\author{Tamir Bendory, Yonina C. Eldar, \emph{Fellow IEEE} and Nicolas Boumal}

\maketitle
{\let\thefootnote\relax\footnotetext{This project received
funding from the European Union's Horizon 2020 research and innovation
program under grant agreement No. 646804-ERCCOG-BNYQ, and from the
Israel Science Foundation under Grant no. 335/14 and {by a research grant from
the Ollendorf Fund}. TB was partially
funded by the Andrew and Erna Finci Viterbi Fellowship. NB is supported by NSF grant DMS-1719558.}} 
\begin{abstract}
The problem of recovering a one-dimensional signal from its Fourier transform magnitude,
called {Fourier} phase retrieval, is ill-posed in most cases. We consider the
closely-related problem of recovering a signal from its phaseless
short-time Fourier transform (STFT) measurements. This problem arises
naturally in several applications, such as ultra-short laser pulse characterization and ptychography. The redundancy offered by the STFT enables unique
recovery under mild conditions. We show that in some cases the unique
solution can be obtained by the principal eigenvector of a matrix,
constructed as the solution of a simple least-squares problem.
When these conditions are not met, we suggest using the principal eigenvector of this matrix to initialize {non-convex local optimization algorithms and propose two such methods. The first is based on minimizing the empirical risk loss function, while the second maximizes a quadratic function on the manifold of phases.} We prove that under appropriate conditions, the proposed initialization is close to the underlying signal. We then analyze the geometry of the empirical risk loss function and show numerically that both gradient algorithms converge to the underlying signal even with small
redundancy in the measurements. In addition, the algorithms are robust
to noise. 
 \end{abstract}



\section{Introduction} \label{sec:introduction}

The problem of recovering a signal from its Fourier transform magnitude
arises in many areas in engineering and science, such as optics, X-ray
crystallography, speech recognition, blind channel estimation, alignment and
astronomy \cite{harrison1993phase,walther1963question,millane1990phase,juang1993fundamentals,baykal2004blind,bendory2017bispectrum, fienup1987phase}.
This problem is called  \emph{{ Fourier} phase retrieval} and can be viewed as
a special case of a quadratic system of equations. The latter area
received considerable attention recently, partially due to its strong
connections with the fields of compressed sensing and matrix completion; 
see for instance \cite{candes2015phase,candes2013phaselift,candes2014phase, shechtman2011sparsity, eldar2014phase,candes2014solving,waldspurger2015phase,waldspurger2016phase}.
Contemporary surveys of the phase retrieval
problem { from a signal processing point of view} can be found in \cite{bendory2017fourier,shechtman2014phase,jaganathan2015phase}.

Phase retrieval for one-dimensional (1D) signals is an ill-posed problem
unless the signal has the minimum phase property \cite{Huang2015,rusu2007extending}. In this special case, the signal can be recovered by several tractable
algorithms (see for instance Section 2.6   of \cite{dumitrescu2007positive}).
Particularly, in \cite{Huang2015} it was shown that a semidefinite
program (SDP) relaxation achieves the optimal solution in the least-squares (LS)  sense. For general signals, two main approaches are typically suggested.
The first builds upon prior knowledge on the signal's support, such
as sparsity or a portion of the underlying signal \cite{fienup1982phase,shechtman2011sparsity,ranieri2013phase,jaganathan2013sparse,shechtman2014gespar,wang2014phase}.
An alternative strategy makes use of additional measurements. Such
measurements can be obtained by structured illuminations and masks
\cite{candes2015phase,candes2014phase,gross2014improved} or by measuring the magnitude
of the short-time Fourier transform (STFT) \cite{eldar2015sparse,jaganathan2015stft}. In \cite{eldar2015sparse},
it was demonstrated that for the same number of measurements,
the STFT magnitude leads to better performance than an over-sampled
discrete Fourier transform~(DFT).

This paper deals with the problem of recovering a 1D signal from its
STFT magnitude. The STFT of a 1D signal $\mathbf{x}\in\mathbb{C}^{N}$
can be interpreted as the Fourier transform of the signal multiplied
by a real sliding window $\mathbf{g}\in \mathbb{R}^N$ with support size $W$ and is defined as 
\begin{equation}
\mathbf{X}[m,k]:=\sum_{n=0}^{N-1}\mathbf{x}[n]\mathbf{g}[mL-n]e^{-2\pi jkn/N},\label{eq:stft}
\end{equation}
where $k=0,\dots,N-1$, $m=0,\dots,\left\lceil \frac{N}{L}\right\rceil -1$
and $L$ determines the separation in time between adjacent sections.
 The pseudo-inverse of the STFT is given by
	\begin{equation} \label{eq:istft}
	\mathbf{x}[n]=\frac{\sum_{m=0}^{\left\lceil \frac{N}{L}\right\rceil-1}\mathbf{\tilde x}[m,n]\overline{\mathbf{g}[mL-n]}}{\sum_{m=0}^{\left\lceil \frac{N}{L} \right\rceil-1} \vert \mathbf{g}[mL-n]\vert^2 },
	\end{equation}
	where $\mathbf{\tilde x}[m,n]$ is the inverse DFT of $\mathbf{X}[m,k]$ for fixed $m$ with respect to the second variable \cite{eldar2015sparse}. 	
In the sequel, all indices should be considered as modulo the signal's
length $N$. We assume that $\mathbf{x}$ and $\mathbf{g}$ are periodically
extended over the boundaries in (\ref{eq:stft}).

The problem of recovering a signal from its STFT magnitude $\vert\mathbf{X}[m,k]\vert^{2}$, frequently called spectrogram, 
arises in several applications in optics and speech processing \cite{nawab1983signal,griffin1984signal}.
Particularly, it serves as the model for a popular variant of an ultra-short
laser pulse characterization technique called Frequency-Resolved Optical
Gating (referred to as X-FROG) \cite{trebino2002frequency,bendory2017uniqueness,bendory2017signal}. Another important 
application is ptychography in which a moving probe is used to sense
multiple diffraction measurements \cite{rodenburg2008ptychography,marchesini2015alternating,maiden2011superresolution,yeh2015experimental}.

Several algorithms were suggested to recover a signal from the magnitude
of its STFT. The classical method, proposed by  Griffin and Lim \cite{griffin1984signal}, is a modification of the alternating projection
(or reduction error) algorithms of Gerchberg and Saxton \cite{gerchberg1972practical}
and Fienup \cite{fienup1982phase}. The properties of this method
are not well understood (for analysis of alternating projection algorithms
in phase retrieval, see \cite{marchesini2015alternating,waldspurger2016phase,pauwels2017fienup}). In \cite{jaganathan2015stft},
the authors prove that a non-vanishing signal can be recovered by
an SDP with maximal overlap between adjacent windows ($L=1$).
They also demonstrate empirically that the algorithm works well with
less restrictive requirements on the window and is robust to noise.
Despite the appealing numerical performance, solving an SDP requires
high computational resources. { Recently, an interesting recovery approach was proposed in \cite{pfander2016robust}.   This paper suggests
a multi-stage method, based on spectral clustering and phase synchronization. It is shown that the algorithm achieves stable estimation (and
exact in the noise-free setting) with only $\mathcal{O}(N\log N)$  phaseless STFT measurements. However, 
this technique requires a random window of length $W=N$, while in most applications it is common to work with shorter windows. Another line of works \cite{iwen2016fast,iwen2016phase} suggest applying a phase synchronization framework \cite{singer2011angular,bandeira2014tightness,boumal2016nonconvex,chen2016projected,perry2016message,bandeira2015non}.
It was shown that even for short windows, the sought signal can be recovered exactly and efficiently by spectral and greedy techniques. These methods are accompanied by stability guarantees.
Their main drawback  is that they rely on reliable estimates of the temporal magnitude, which do not always exist.}

Here, we take a different approach and propose a data-driven initialization technique, followed by non-convex gradient algorithms.
We begin by taking the 1D DFT of the acquired data
with respect to the frequency variable (the second variable of the
STFT). This transformation reveals the underlying structure of the
data and greatly simplifies the analysis. As a direct consequence,
we show that for $L=1$ and sufficiently long windows $W\geq \left\lceil\frac{N+1}{2}\right\rceil$
(and some mild additional conditions), one can recover the signal
by extracting the principal eigenvector of a designed matrix, constructed as the  solution of a simple linear LS problem. We refer to this matrix as the  \emph{approximation matrix} since it approximates the correlation matrix $\mathbf{X}:=\mathbf{x}\mathbf{x}^*$. 
 
 When the conditions for a closed-form solution are not met, 
 we propose using the principal eigenvector of the approximation matrix to initialize two non-convex algorithms. The first  is based on minimizing a standard quadratic loss function, frequently called the empirical risk (ER). Inspired by the phasecut method  \cite{goemans1995improved,waldspurger2015phase}, we also propose a  new phase retrieval algorithm, called Non-Convex PhaseCut (NCPC), that maximizes a quadratic function over the set of phases. Each step of the algorithm follows the component of the gradient which agrees with the phase constraints.
 	 As will be shown, the ER technique is more stable in the low signal--to--noise ratio (SNR) regimes, while NCPC is superior in high SNR environments and for short windows. Our approach deviates in two important aspects from the recent line of work  in non-convex phase retrieval \cite{candes2015Wirtinger,chen2015solving,netrapalli2015phase,sanghavi2015local,wang2016solving,waldspurger2016phase,zhang2016reshaped}.  First, all these papers focus their attention on the setup of phase retrieval with random sensing vectors and rely heavily on {probabilistic} considerations.  In this case,  efficient algorithms were designed to estimate the signal from $\mathcal{O}(N)$ measurements. In contrast, we consider a deterministic framework.   
Second, we construct our approximation matrix by the solution of a LS problem, whereas the aforementioned papers take a superposition of the measurements to approximate $\mathbf{X}$.

The properties of { non-convex  algorithms} depend heavily on the initialization method and the geometry of the loss functions.
For $L=1$, we estimate
the distance between the proposed initialization and the target signal,
which decays to zero as $W$ tends to $\frac{N+1}{2}$. If the signal has unit
modulus entries, then a slight modification of our initialization recovers the signal
exactly for $W\geq2$. In the later case, we also prove the existence
of a \emph{basin of attraction} around the global minimum of the ER loss
function and estimate its size. In the basin of attraction, the
algorithm is guaranteed to converge to a global minimum at a geometric 
rate. We note that while the theoretical guarantees
of the algorithms are limited, their experimental performance is significantly
better. Particularly, the algorithms perform well with small
redundancy in the measurements and are robust in the presence
of noise.

The paper is organized as follows. We begin in Section \ref{sec:Problem-Formulation}
by formulating mathematically the problem of phase retrieval from
STFT magnitude measurements. In Section \ref{sec:uniqueness} we discuss
the uniqueness of the solution and present conditions under which it
has a closed-form LS expression. Additionally, we present a method that recovers signals with unit modulus entries under mild conditions. 
Section \ref{sec:Algorithm} presents
the { two non-convex  algorithms} with the proposed initialization. Section \ref{sec:Numerical-Results}
shows numerical results and Section \ref{sec:theory} presents our
theoretical findings regarding the proposed initialization and the ER loss function.  Proofs
are provided in Section \ref{sec:Geometry}. Section
\ref{sec:Conclusion} concludes the paper, discusses its main implications
and draws potential future research directions.

Throughout the paper we use the following notation. Boldface small and
capital letters denote vectors and matrices, respectively. We use
$\mathbf{Z}^{T}$ and $\mathbf{Z}^{*}$ for the transpose and Hermitian
of a matrix $\mathbf{Z}$; similar notation is used for vectors. We
further use $\mathbf{Z}^{\dagger}$ and tr$\left(\mathbf{Z}\right)$
for the Moore\textendash Penrose pseudo-inverse and the trace of the
matrix $\mathbf{Z}$, respectively. The $\ell${th} circular diagonal of a matrix
$\mathbf{Z}$ is denoted by $\diag\left(\mathbf{Z},\ell\right)$. Namely,
$\diag\left(\mathbf{Z},\ell\right)$ is a column vector with entries
$\mathbf{Z}\left[i,\left(i+\ell\right)\bmod N\right]$ for $i=0,\dots,N-1$.
We define the sign of a complex number $a$ as $\operatorname{phase}\left(a\right):=\frac{a}{\left|a\right|}$ for $a\neq 0$ and zero otherwise. We also  use $'\odot'$, $'\circ'$ and $'\ast'$ for the Hadamard (point-wise) product, composition of functions
and convolution, respectively.  The set of all complex (real) signals
of length $N$ whose entries have modulus \textbf{$a>0$} are denoted by $\mathbb{C}_{a}^{N}$
($\mathbb{R}_{a}^{N}$). Namely, $\mathbf{z}\in\mathbb{C}_{a}^{N}$
means that $\left|\mathbf{z}[n]\right|=a$ for all $n$.


\section{\label{sec:Problem-Formulation}Problem Formulation}

We aim at recovering an underlying signal $\mathbf{x}\in\mathbb{C}^N$ from the
magnitude of its STFT, i.e., from measurements 
\begin{equation}
\mathbf{Z}[m,k]=\left|\mathbf{X}[m,k]\right|^{2}.\label{eq:meas0}
\end{equation}
Note that the signals $\mathbf{x}$ and $\mathbf{x}e^{j\phi}$ yield
the same measurements for any \emph{global phase} $\phi\in\mathbb{R}$
and therefore the phase $\phi$ cannot be recovered by any method.
This \emph{global phase ambiguity} leads naturally to the following
definition: 
\begin{defn}
\label{def:dis}The distance between two vectors is defined as 
\[
d\left(\mathbf{z},\mathbf{x}\right)=\min_{\phi\in[0,2\pi)}\left\Vert \mathbf{z}-\mathbf{x}e^{j\phi}\right\Vert _{2}.
\]
If $d\left(\mathbf{z},\mathbf{x}\right)=0$ then we say that $\mathbf{x}$
and $\mathbf{z}$ are equal \emph{up to global phase}. The phase $\phi\in[0,2\pi)$
attaining the minimum is denoted by $\phi(\mathbf{z})$, i.e., 
\[
\phi(\mathbf{z})=\arg\min_{\phi\in[0,2\pi)}\left\Vert \mathbf{z}-\mathbf{x}e^{j\phi}\right\Vert _{2}.
\]
\end{defn}
{ In the sequel, we make use of the notion of \emph{non-vanishing signals}, defined as follows:
\begin{defn} \label{def:non_vanishing}
A vector $\mathbf{z}\in\mathbb{C}^N$ is called non-vanishing  if $\mathbf{z}[n]\neq 0 $ for all $n=0,\dots,N-1$.
\end{defn}
}

Instead of treating the measurements (\ref{eq:meas0}) directly, we often 
consider the acquired data in a transformed domain by taking its 1D
DFT with respect to the frequency variable (normalized by $1/N$).
Then, our measurement model reads 
\begin{equation}
\begin{split} & \mathbf{Y}[m,\ell]=\frac{1}{N}\sum_{k=0}^{N-1}\mathbf{Z}[m,k]e^{-2\pi jk\ell/N}\\
 & =\sum_{n=0}^{N-1}\mathbf{x}[n]\overline{\mathbf{x}[n+\ell]}\mathbf{g}[mL-n]\overline{\mathbf{g}[mL-n-\ell]}.\label{eq:meas}
\end{split}
\end{equation}
When $W\leq \ell\leq N-W $, we have $\mathbf{Y}[m,\ell]=0$ for all $m$. In this sense, $\mathbf{Y}[m,\ell]$ can be interpreted as a ``$W$ -- bandlimited'' function.
 Observe that for fixed $m$, $\mathbf{Y}[m,\ell]$
is simply the auto-correlation of $\mathbf{x}\odot\mathbf{g}_{mL}$,
where $\mathbf{g}_{mL}:=\left\{ \mathbf{g}[mL-n]\right\} _{n=0}^{N-1}$.

We will make repetitive use of several representations of the data.
The first is based on a matrix formulation. Let $\mathbf{D}_{mL}\in\mathbb{R}^{N\times N}$
be a diagonal matrix composed of the entries of $\mathbf{g}_{mL}$.
Let $\mathbf{P}_{\ell}$ be a matrix that shifts (circularly) the
entries of a vector by $\ell$ locations, namely, $\left(\mathbf{P}_{\ell}\mathbf{x}\right)\left[n\right]=\mathbf{x}\left[n+\ell\right]$.
Then, the correlation matrix $\mathbf{X}:=\mathbf{x}\mathbf{x}^{*}$
is mapped linearly to $\mathbf{Y}[m,\ell]$ as follows: 
\begin{eqnarray}
\mathbf{Y}[m,\ell] & = & \left(\mathbf{D}_{mL-\ell}\mathbf{D}_{mL}\mathbf{P}_{\ell}\mathbf{x}\right)^{*}\mathbf{x}\nonumber \\
 & = & \mathbf{x}^{*}\mathbf{H}_{m,\ell}\mathbf{x}\nonumber \\
 & = & \mbox{tr}\left(\mathbf{X}\mathbf{H}_{m,\ell}\right),\label{eq:matrix}
\end{eqnarray}
where 
\begin{equation}
\mathbf{H}_{m,\ell}:=\mathbf{P}_{-\ell}\mathbf{D}_{mL}\mathbf{D}_{mL-\ell}.\label{eq:H}
\end{equation}
Observe that $\mathbf{P}_{\ell}^{T}=\mathbf{P}_{-\ell}$ and 
$\mathbf{H}_{m,\ell}=0$ for $W\leq \ell\leq N-W $. 
{Similarly, the STFT magnitude in (\ref{eq:meas0}) (before the 1D DFT) can be written as 
\begin{equation} \label{eq:matrix_Z}
\mathbf{Z}[m,k]=\mathbf{x}^{*}\mathbf{\widetilde{H}}_{m,k}\mathbf{x},
\end{equation}
where
\begin{equation}
\mathbf{\widetilde{H}}_{m,k}:=   \mathbf{D}_{mL}\mathbf{f}_k^{}\mathbf{f}_k^*\mathbf{D}_{mL}, \label{eq:H_tilde}
\end{equation}
and $\mathbf{f}_k^*$ is the $k$th row of the DFT matrix.}

An alternative useful representation of (\ref{eq:meas}) is as multiple
systems of linear equations. For fixed $\ell\in \{-(W-1),\dots,W-1\}$ we have
\begin{equation}
\mathbf{y}_{\ell}=\mathbf{G}_{\ell}\mathbf{x}_{\ell},\label{eq:Gl}
\end{equation}
where $\mathbf{y}_{\ell}:=\left\{ \mathbf{Y}[m,\ell]\right\} _{m=0}^{\frac{N}{L}-1}$
and $\mathbf{x}_{\ell}:=\diag\left(\mathbf{X},\ell\right)$.
The $(m,n)$th entry of the matrix $\mathbf{G}_{\ell}\in\mathbb{R}^{\left\lceil\frac{N}{L}\right\rceil\times N}$
is given by $\mathbf{g}[mL-n]\mathbf{g}[mL-n-\ell]$. For $L=1$, $\mathbf{G}_{\ell}$ is a circulant matrix. We recall that
a circulant matrix is diagonalized by the DFT matrix, namely, it can be 
factored as $\mathbf{G}_{\ell}=\mathbf{F}^{-1}\mathbf{\Sigma}_{\ell}\mathbf{F},$
where $\mathbf{F}$ is the DFT matrix and $\mathbf{\Sigma}_{\ell}$ is a diagonal
matrix, whose entries are given by the DFT of the first column of $\mathbf{G}_{\ell}$.
In this case, the first column is given by $\mathbf{g}\odot\left(\mathbf{P}_{-\ell}\mathbf{g}\right)$. Therefore the matrix $\mathbf{G}_{\ell}$ is invertible if and
only if the DFT of $\mathbf{g}\odot\left(\mathbf{P}_{-\ell}\mathbf{g}\right)$
is non-vanishing.

Our problem of recovering $\mathbf{x}$ from the measurements (\ref{eq:meas0}) can therefore be posed as a  constrained LS problem: 
 \begin{eqnarray}
 &  & \min_{\mathbf{\tilde{X}}\in\mathcal{H}^{N}}\sum_{\ell=-\left(W-1\right)}^{W-1}\left\Vert \mathbf{y}_{\ell}-\mathbf{G}_{\ell}\diag\left(\mathbf{\tilde{X}},\ell\right)\right\Vert _{2}^{2}\nonumber \\
 &  & \mbox{subject to}\quad\mathbf{\tilde{X}}\succeq 0,\quad \mbox{rank}\left(\tilde{\mathbf{X}}\right)=1, \label{eq:LS_rank-2}
\end{eqnarray}
where $\mathcal{H}^{N}$ is the set of all Hermitian matrices of size $N\times N$.
In the spirit of \cite{goemans1995improved,waldspurger2015phase,candes2015phase,shechtman2011sparsity}, STFT phase retrieval may then be relaxed to a tractable SDP by dropping the rank constraint. 
{ In the noiseless case, this SDP relaxation is equivalent to the one suggested in \cite{jaganathan2015stft} since the conditions on $\tilde{\mathbf{X}}$ to achieve zero objective function are the same, up to a Fourier transformation.} While the SDP relaxation technique has shown good numerical performance for the recovery from phaseless STFT measurements, it requires solving the problem in a lifted domain with $N^2$ variables. We take a different route to reduce the computational load. In the next section, we show that (\ref{eq:LS_rank-2}) admits a unique solution under moderate conditions. We further show that it has a  
 closed-form LS solution when the window $\mathbf{g}$ is sufficiently long. If the conditions for the LS solution are not met, then we suggest   { two non-convex algorithms}. To initialize the { algorithms}, we approximate (\ref{eq:LS_rank-2}) in two stages by first solving the LS objective function and then extracting its principal eigenvector.


\section{\label{sec:uniqueness} Uniqueness and Basic Algorithms}

A fundamental question in phase retrieval problems is whether the quadratic measurement operator of (\ref{eq:meas0}), or equivalently the non-convex problem (\ref{eq:LS_rank-2}), 
determines the underlying signal $\mathbf{x}$ uniquely (up to global
phase, see Definition \ref{def:dis}). In other words, one wants to know the 
conditions on the window $\mathbf{g}$ and the signal $\mathbf{x}$ such that the non-linear transformation
that maps $\mathbf{x}$ to $\mathbf{Z}$ is injective. Before treating this question, we 
introduce some basic window definitions:
\begin{defn}
A window $\mathbf{g}$ is called a \emph{rectangular window of length
$W$ }if $\mathbf{g}[n]=1$ for all $n=0,\dots,W-1$ and zero elsewhere.
It is a \emph{non-vanishing window of length $W$ }if $\mathbf{g}[n]\neq0$
for all $n=0,\dots,W-1$ and zero elsewhere. 
\end{defn}
According to (\ref{eq:LS_rank-2}), the injectivity of the measurement operator
is related to the window's length $W$ and the invertibility of the matrices $\mathbf{G}_{\ell}$ for $\vert \ell \vert<W$. For that reason, 
we give special attention to windows for which
the associated matrices are invertible. 
\begin{defn}
\label{def:admissible_win}A window $\mathbf{g}$ is called an \emph{admissible
window of length $W$} if for all $\ell=-(W-1),\dots,W-1$ the following
two equivalent properties hold: 
\begin{enumerate}
\item The DFT of the vector $\mathbf{g}\odot\left(\mathbf{P}_{-\ell}\mathbf{g}\right)$
is non-vanishing. 
\item The associated circulant matrices $\mathbf{G}_{\ell}$ as given in
(\ref{eq:Gl}) are invertible. 
\end{enumerate}
\end{defn}
{ An important example for an admissible window is a rectangular window. Specifically, we have the following lemma:}
\begin{lem}
\label{claim:rect_win}A rectangular window $\mathbf{g}$ of length $2\leq W\leq N/2$ is an admissible window of length $W$
if $\alpha$ and $N$ are co-prime numbers for all $\alpha=2,\dots,W$. This holds trivially when $N$ is a prime
number. \end{lem}
\begin{IEEEproof}
Observe that $\mathbf{g}\odot\left(\mathbf{P}_{-\ell}\mathbf{g}\right)$
is a rectangular window of length $W-\left|\ell\right|$ for $\ell=-(W-1)\dots,W-1$.
The DFT of a rectangular window of size $W-\left|\ell\right|$ is
a Dirichlet kernel which is non-vanishing if $W-\left|\ell\right|$
and $N$ are co-prime. 
\end{IEEEproof}
{ 
The family of admissible windows contains more examples.
To demonstrate this, we consider a non-vanishing window of length $W$ whose entries are i.i.d.\ normal variables. We then compute the minimal  absolute value of the DFT of $\mathbf{g}\odot\left(\mathbf{P}_{-\ell}\mathbf{g}\right)$ for all $\ell=-(W-1),\dots,W-1$, namely,
\begin{equation} \label{eq:min_ev}
\vert\lambda_{\min}\vert = \min_{k,\vert \ell\vert\leq W }\vert \left( \mathbf{F}\left(\mathbf{g}\odot\left(\mathbf{P}_{-\ell}\mathbf{g}\right)\right)\right)[k]\vert.
\end{equation}
 We repeated this process 100 times for several values of $W$. As can be seen in Table \ref{tab:1}, $\vert\lambda_{\min}\vert$ is bounded away from zero, implying that the windows are indeed admissible.

\begin{table*}
\begin{centering}
\begin{tabular}{|c|c|c|c|c|}
\hline 
\backslashbox{$\vert \lambda_{\min}\vert $ }{ Window's length}
  & $W=5$ & $W=10$ & $W=15$ & $W=20$\tabularnewline
\hline 
\hline 
Mean  &     0.0463  &     0.0367
  & 0.0426
  & 0.0549 \tabularnewline
\hline 
Min & 0.0008 & 0.0021 & 0.0019 & 0.0031\tabularnewline
\hline 
\end{tabular}
\par\end{centering}

\protect\caption{\label{tab:1} The mean and minimal values of $\vert \lambda_{\min}\vert $ for windows of length $W$ with i.i.d.\ normal entries as defined in (\ref{eq:min_ev}) over 100 experiments for different window lengths and $N=25$.}
\end{table*} 
}

We now analyze the uniqueness of the measurement operator for the case $L=1$. Uniqueness results for  $L>1$ are discussed in  \cite{nawab1983signal,jaganathan2015stft}. Our  results are constructive in the sense that their proofs
provide an explicit scheme to recover the signal.

Our first  result concerns non-vanishing signals. In this case, the magnitude of the STFT determines the
underlying signal uniquely under mild conditions. This conclusion was
already derived in \cite{bojarovska2015phase} based on different
considerations. Nevertheless, the following proposition comes with an explicit recovery scheme as presented in Appendix  \ref{sec:algebriac_alg}.
\begin{proposition}
\label{thm:algebriac}Let $L=1$. Suppose that $\mathbf{x}$ is non-vanishing
and that the DFT of $\mathbf{g}\odot\left(\mathbf{P}_{-\ell}\mathbf{g}\right)$
is non-vanishing for $\ell=0,1$. Then, $\left|\mathbf{X}[m,k]\right|^{2}$
determines $\mathbf{x}$ uniquely (up to global phase).\end{proposition}
\begin{IEEEproof}
See Appendix \ref{sec:algebriac_alg}. 
\end{IEEEproof}
A similar uniqueness result was derived in \cite{eldar2015sparse}.
There, it is required that the DFT of $\vert\mathbf{g}[n]\vert^{2}$
is non-vanishing, $N\geq2W-1$ and $N$ and $W-1$ are co-prime numbers.

In the special case in which the signal is known to have unit modulus
entries, the signal can be recovered as the principal eigenvector of a matrix  designed  as follows:

\begin{proposition} \label{cor:init_unit_signal} Let  $L=1$. Suppose that $\mathbf{x}\in\mathbb{C}_{1/\sqrt{N}}^{N}$
and that $\mathbf{g}$ is an admissible window of length $W\geq 2$. Fix $M\in \{1,\dots,W-1\}$ and
let $\mathbf{X}_{0}$ be a matrix defined by
\begin{equation}
\diag\left(\mathbf{X}_{0},\ell\right)=\begin{cases}
\mathbf{G}_\ell^{-1}\mathbf{y}_\ell, & \quad\ell=0,M,\\
0, & \quad \mbox{otherwise},
\end{cases}\label{eq:matrix_construction}
\end{equation}
where $\mathbf{G_\ell}$ and $\mathbf{y_\ell}$ are defined in (\ref{eq:Gl}).
Then, $\mathbf{x}$ (up to global phase) is a principal eigenvector of $\mathbf{X}_0$.
 \end{proposition} 
\begin{IEEEproof}
See Appendix \ref{proof_of_prop_unit_signal}. 
\end{IEEEproof}

For general signals (not necessarily non-vanishing) and $L=1$, we next derive a LS algorithm that stably recovers any complex signal  if the window is sufficiently long. In the absence of noise, the recovery is exact (up to global phase). The method, summarized in Algorithm \ref{alg:linearLS}, is based on constructing a matrix $\mathbf{X}_0$ that approximates  the correlation matrix
$\mathbf{X}:=\mathbf{x}\mathbf{x}^{*}$.
The $\ell${th} diagonal of $\mathbf{X}_0$  is chosen  as the  solution of the LS problem $\min_{\tilde{\mathbf{x}}\in\mathbb{C}^N}\|\mathbf{y}_\ell-\mathbf{G}_\ell\mathbf{\tilde{x}}\|_2$ (see (\ref{eq:Gl})). If the matrix $\mathbf{G}_{\ell}$ is invertible, then  
\[\diag\left(\mathbf{X}_0,\ell\right)=\mathbf{G}_\ell^{-1}\mathbf{y}_\ell=\diag\left(\mathbf{X},\ell\right).
\]
 Therefore, when all matrices  $\mathbf{G}_\ell$ are invertible,  $\mathbf{X}_0=\mathbf{X}$. In order to estimate $\mathbf{x}$, the (unit-norm) principal eigenvector of $\mathbf{X}_0$ is normalized by
 \begin{equation}\label{eq:alpha}
\alpha=\sqrt{\sum_{n\in P}\left(\mathbf{G}_{0}^{\dagger}\mathbf{y}_{0}\right)[n]}, 
 \end{equation}
where $P:=\{ n\thinspace:\thinspace(\mathbf{G}_{0}^{\dagger}\mathbf{y}_{0})[n]>0\} $. If $\mathbf{G}_0$ is invertible then 
\[ \sum_{n= 0}^{N-1}\left(\mathbf{G}_{0}^{-1}\mathbf{y}_{0}\right)[n]=\sum_{n= 0}^{N-1}\left(\diag\left(\mathbf{X},0\right)\right)[n]= \|\mathbf{x}\|_2^2=\lambda_0,\]where $\lambda_0$ is the top eigenvalue of $\mathbf{X}$. 
If $\mathbf{G}_{0}$ is not invertible or in the presence of noise,
some terms of the vector $\mathbf{G}_{0}^{\dagger}\mathbf{y}_0$
might be negative. In this case, we estimate $\|\mathbf{x}\|_{2}$ by summing only the positive terms (the set $P$ in (\ref{eq:alpha})).
Note that all matrix inversions can
be performed efficiently using the FFT due to the circulant structure of
$\mathbf{G}_{\ell}$.

 The following proposition shows that  Algorithm \ref{alg:linearLS} recovers
the underlying signal for $L=1$ if the window is sufficiently long and satisfies some additional technical conditions.
 In \cite{bojarovska2015phase}, an equivalent uniqueness result was derived
but without providing an algorithm. Algorithm \ref{alg:linearLS} 
is equivalent to the discretized version of Wigner deconvolution that was suggested previously without
theoretical analysis in \cite{rodenburg1992theory,yang2011iterative}.
\begin{proposition}
\label{thm:linearLS}Let $L=1$ and suppose that $\mathbf{g}$ is
an admissible window of length $W\geq\left \lceil\frac{N+1}{2}\right\rceil$ (see Definition \ref{def:admissible_win}).
Then, Algorithm \ref{alg:linearLS} recovers any complex
signal uniquely up to global phase.\end{proposition}
\begin{IEEEproof}
See Appendix \ref{sec:linearLS}. 
\end{IEEEproof}

\begin{algorithm}
\textbf{Input:} The measurements $\mathbf{Z}[m,k]$ as given in (\ref{eq:meas0}).\\
\textbf{Output:} $\mathbf{x}_{0}$: estimation of \textbf{$\mathbf{x}$}. 
\begin{enumerate}
\item Compute $\mathbf{Y}\left[m,\ell\right]$, the 1D DFT with respect to the second variable of $\mathbf{Z}[m,k]$
as given in (\ref{eq:meas}).

\item Construct a matrix $\mathbf{X}_{0}$ such that 
\[
\diag\left(\mathbf{X}_{0},\ell\right)=\begin{cases}
\mathbf{G}_{\ell}^{\dagger}\mathbf{y}_{\ell} & \ell=-\left(W-1\right),\cdots,\left(W-1\right),\\
0 & \mbox{otherwise,}
\end{cases}
\]
where $\mathbf{G}_{\ell}\in\mathbb{R}^{N\times N}$ are defined  in (\ref{eq:Gl}). 
\item Let $\mathbf{x}_{p}$ be the principal (unit-norm) eigenvector of
$\mathbf{X}_{0}$. Then, 
\[
\mathbf{x}_{0}=\sqrt{\sum_{n\in P}\left(\mathbf{G}_{0}^{\dagger}\mathbf{y}_{0}\right)[n]}\mathbf{x}_{p},
\]
where $P:=\left\{ n\thinspace:\thinspace\left(\mathbf{G}_{0}^{\dagger}\mathbf{y}_{0}\right)[n]>0\right\} $. 
\end{enumerate}
\protect\caption{\label{alg:linearLS}Least-squares algorithm for $L=1$}
\end{algorithm}

In many cases, the window is shorter than $\left\lceil\frac{N+1}{2}\right\rceil$ so that~\eqref{eq:LS_rank-2} may not admit a closed-form
LS solution. In these cases, we { propose two non-convex recovery algorithms. The first is a standard ER minimization that seems to work well in low SNR regimes. The second maximizes a quadratic function over the manifold of phases. This approach shows superior performance for short windows and high SNR.  
 }
In order to initialize these algorithms, we use the same LS-based method of Algorithm~\ref{alg:linearLS}. However, for short windows we cannot estimate $\diag\left(\mathbf{X},\ell\right)$ for $\ell=W,\dots,N-W$ as the matrices  $\mathbf{G}_\ell$ are simply zero.
 Nonetheless, we will show  by both theoretical results and numerical experiments that under appropriate conditions, the principal eigenvector of the approximation matrix $\mathbf{X}_0$, with appropriate normalization,     
is a good initial estimator 
of $\mathbf{x}$.


\section{\label{sec:Algorithm} Local Non-Convex Algorithms }

In this section we present our main algorithmic approach to recover
a signal from its STFT magnitude (\ref{eq:meas0}). {First, we propose two non-convex gradient algorithms  to estimate the signal. 
	  As the problem is inherently non-convex, we then suggest a systematic, data--driven, technique for initialization.  This non-convex approach for STFT phase retrieval is summarized in Algorithm~\ref{alg:non_convex}.  
	  The code for all algorithms is publicly available at 
	  \url{ http://webee.technion.ac.il/Sites/People/YoninaEldar}.

	
\subsection{Empirical Risk Minimization} \label{sec:ER}

	Recall that  the STFT magnitude can be written as $\mathbf{Z}[m,k]=\mathbf{x}^{*}\mathbf{\widetilde{H}}_{m,k}\mathbf{x}$, where $\mathbf{\widetilde{H}}$ is given in (\ref{eq:H_tilde}). Alternatively, 
 by taking the 1D DFT with respect to the frequency variable, the measurement
model becomes $\mathbf{Y}[m,\ell]=\mathbf{x}^{*}\mathbf{H}_{m,\ell}\mathbf{x}$,
where $\mathbf{H}_{m,\ell}$ is defined in (\ref{eq:H}). It is therefore
natural to minimize the empirical risk (ER) loss function: 
\begin{align}
f(\mathbf{u})&=\frac{1}{2}\sum_{m=0}^{\left\lceil N/L\right\rceil-1}\sum_{k=0}^{N-1}\left\vert\mathbf{u}^{*}\mathbf{\widetilde{H}}_{m,k}\mathbf{u}-\mathbf{Z}[m,k]\right\vert^{2} \label{eq:LS_Z} \\
&=\frac{1}{2}\sum_{m=0}^{\left\lceil N/L\right\rceil-1}\sum_{\ell=-\left(W-1\right)}^{W-1}\left\vert \mathbf{u}^{*}\mathbf{H}_{m,\ell}\mathbf{u}-\mathbf{Y}[m,\ell]\right\vert^{2}.
\label{eq:LS}
\end{align}
The equality between the two loss functions is proven in Appendix \ref{sec:proof_of_equivalence}. 
In the sequel, we use both formulations.}

\begin{figure}
\centering
\includegraphics[scale=0.6]{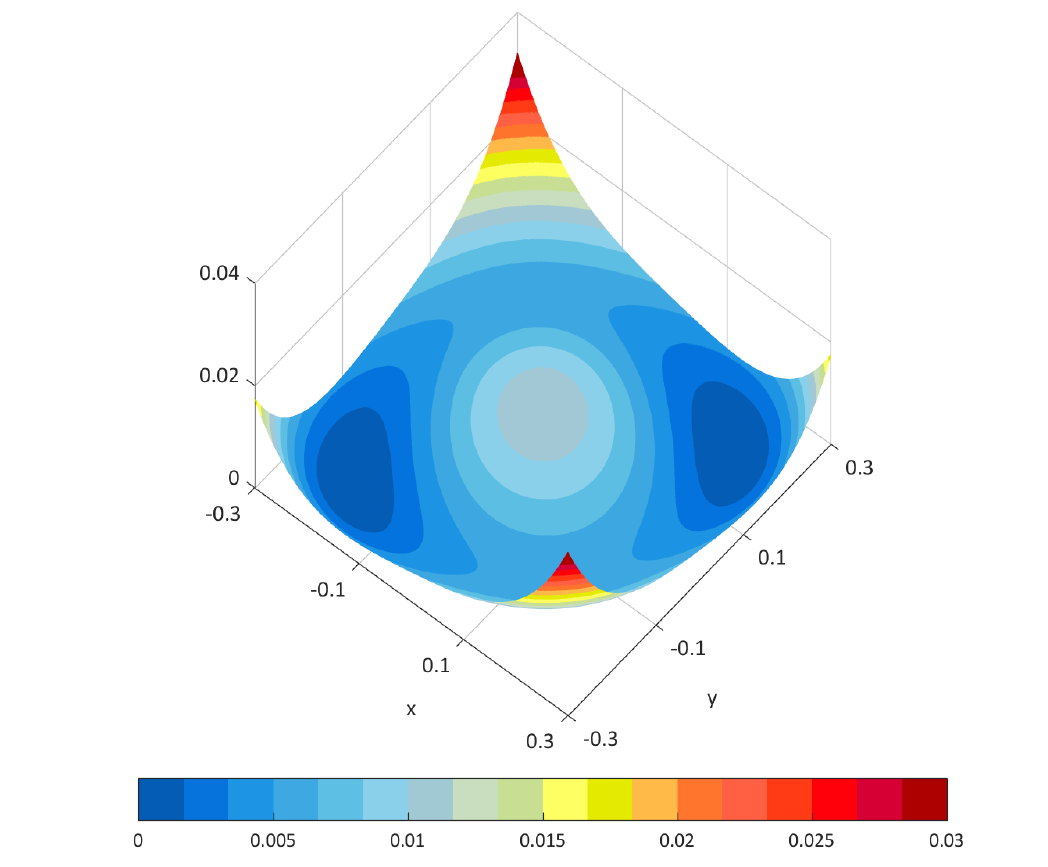}
\caption{\label{fig:loss_function} The two-dimensional (first two variables) plane of the loss function
(\ref{eq:LS}) of the signal $\mathbf{x}=[0.2,0.2,0,0,0]$ (i.e., $N=5$)
with $L=1$ and a rectangular window of length $W=2$.}
\end{figure}

Figure \ref{fig:loss_function} presents the two-dimensional (first two variables) plane of the loss
function (\ref{eq:LS_Z}) for the signal $\mathbf{x}=[0.2,0.2,0,0,0]$
(i.e., $N=5$) with $L=1$ and a rectangular window of length $W=2$.
The function has no sharp transitions and contains two saddle points and two
global minima (as a result of the global phase ambiguity). Accordingly,
in this specific case and bearing in mind that our view is restricted to two of the five dimensions only, it seems that a gradient descent algorithm  will converge to a global minimum from
almost any  initialization (see also \cite{lee2016gradient}). While this 
 phenomenon does not occur for any arbitrary parameter selection, this example 
motivates applying a gradient algorithm directly on the  non-convex loss function (for a similar demonstration of the loss function
with random sensing vectors, see \cite{sun2016geometric}).

One way to minimize the ER loss function~\eqref{eq:LS_Z} or~\eqref{eq:LS}  is by employing a gradient algorithm, where the $k${th} iteration takes on the form 
\[
\mathbf{x}_{k}=\mathbf{x}_{k-1}-\mu\nabla f\left(\mathbf{x}_{k-1}\right),
\]
for step size $\mu$. 
{ For real signals,} direct computation of the gradient in~\eqref{eq:LS_Z} gives  
\begin{align}
\nabla f(\mathbf{u}) & =\sum_{m=0}^{\left\lceil N/L\right\rceil-1}\sum_{k=0}^{N-1}\left(h(\mathbf{u})-\mathbf{Z}[m,k]\right)\nabla{h}(\mathbf{u}),\label{eq:grad}
\end{align}
where 
\begin{align*}
h(\mathbf{u}) & :=\mathbf{u}^{T}\widetilde{\mathbf{H}}_{m,k}\mathbf{u},\quad  \nabla{h}(\mathbf{u})  =2\widetilde{\mathbf{H}}_{m,k}\mathbf{u}.
\end{align*}
Similar computations can be performed for~\eqref{eq:LS}.
{If the signal is complex, then one can use the elegant formulation of Wirtinger derivatives, see \cite{kreutz2009complex,candes2015Wirtinger,sun2016geometric}. The loss functions~\eqref{eq:LS_Z} or~\eqref{eq:LS} may be minimized by many other methods. For instance, in Section \ref{sec:Numerical-Results} we employ a trust-region algorithm.}

\begin{algorithm}
\textbf{Input:} The measurements $\mathbf{Z}[m,k]$ as given in (\ref{eq:meas0}).\\
\textbf{Output:} $\hat{\mathbf{x}}$: estimation of \textbf{$\mathbf{x}$}. 
\begin{enumerate}
\item \textbf{Initialization}: Apply Algorithm \ref{alg:linearLS} (for $L=1$) or Algorithm \ref{alg:linearLS_L} (for $L>1)$.
\item \textbf{Refinement}: Use the output of stage 1 to initialize a gradient algorithm that minimizes the empirical risk (Section \ref{sec:ER}) or the Non-Convex PhaseCut (NCPC) of Algorithm \ref{alg:PR}.
\end{enumerate}
\protect\caption{\label{alg:non_convex} Non-convex approach for STFT phase retrieval}
\end{algorithm}


\subsection{Non-Convex PhaseCut (NCPC)} \label{sec:ncpc}


\subsubsection{The Algorithm}
When  minimizing the empirical risk~\eqref{eq:LS_Z} or~\eqref{eq:LS}, the unknown signal itself is the optimization variable. Alternatively, we may take the point of view that the unknowns are the phases of the STFT measurements. Indeed, if these phases were known, then one could recover the signal by applying~\eqref{eq:istft}. We may therefore rework the problem into one where only the phases are variables~\cite{waldspurger2015phase}. 

Thus, we aim to estimate $\mathbf{U} \in \mathbb{C}^{\left\lceil\frac{N}{L}\right\rceil \times N}$ with unit-modulus entries such that $\mathbf{X} \approx \mathbf{Z}^{1/2} \odot \mathbf{U}$, that is, we wish to recover the missing phases. One may propose to estimate these and the signal $\mathbf{{x}}$ simultaneously by minimizing $\|\mathbf{Z}^{1/2} \odot \mathbf{U} - \STFT(\mathbf{x})\|_{\mathrm{F}}^2$ over both $\mathbf{x}$ and $\mathbf{U}$, where $\STFT(\mathbf{x})$ maps $\mathbf{x}$ to its STFT following~\eqref{eq:stft}. Assuming $\mathbf{U}$ is fixed, the  solution for $\mathbf{x}$ is $\mathbf{x} = \STFTdagger(\mathbf{Z}^{1/2} \odot \mathbf{U})$, where  the operator  $\STFTdagger$  is given by~\eqref{eq:istft}.
By substitution, we obtain an optimization problem in terms of $\mathbf{U}$ only:
\begin{align*}
\min_{\mathbf{U} \in \mathbb{C}^{ \left\lceil \frac{N}{L}\right\rceil \times N}} & \|(\mathbf{I} - \STFT \circ \STFTdagger) (\mathbf{Z}^{1/2} \odot \mathbf{U})\|_{\mathrm{F}}^2 \\ \textrm{ subject to } & \quad |\mathbf{U}[m,k]| = 1,\thinspace \forall m, k.
\end{align*}
Since $\mathbf{I} - \STFT \circ \STFTdagger$ is an orthogonal projector, this further simplifies into the following non-convex optimization problem over complex phases:
\begin{align*}
\min_{\mathbf{U} \in \mathbb{C}^{ \left\lceil \frac{N}{L}\right\rceil \times N}} & \left\langle \mathbf{Z}^{1/2} \odot \mathbf{U}, (\mathbf{I}  - \STFT \circ \STFTdagger) (\mathbf{Z}^{1/2} \odot \mathbf{U}) \right\rangle \\ \textrm{ subject to } & \quad |\mathbf{U}[m,k]| = 1,\thinspace \forall m, k,
\end{align*}
where we use the Frobenius inner product
\begin{align}
	\left\langle \mathbf{A}, \mathbf{B} \right\rangle & = \Re\left\{ \operatorname{Trace}\left(\mathbf{A}^* \mathbf{B} \right) \right\}.
	\label{eq:frobinner}
\end{align}
The term involving the identity operator $\mathbf{I}$ is constant under the constraints, so that the problem is equivalent to the following maximization problem:
\begin{align}
	\max_{\mathbf{U} \in \mathbb{C}^{ \left\lceil \frac{N}{L}\right\rceil \times N}} & \left\langle \mathbf{Z}^{1/2} \odot \mathbf{U}, \STFT \circ \STFTdagger (\mathbf{Z}^{1/2} \odot \mathbf{U}) \right\rangle \nonumber \\ \textrm{ subject to } & \quad |\mathbf{U}[m,k]| = 1,\thinspace \forall m, k.
\label{eq:PRoptim}
\end{align}

Notice that $ \STFT \circ \STFTdagger$ is the orthogonal projector onto the subspace of matrices which are the STFT of some signal. As a result, applying $ \STFT \circ \STFTdagger$ to the matrix $\mathbf{Z}^{1/2} \odot \mathbf{U}$ produces the matrix which, in the LS sense, is closest to being the STFT of a signal. Thus, the cost function in~\eqref{eq:PRoptim} favors phases $\mathbf{U}$ such that $\mathbf{Z}^{1/2} \odot \mathbf{U}$ is as close as possible to an  STFT. We recall  that this projection operator can be computed efficiently by applying~\eqref{eq:stft} and~\eqref{eq:istft} using FFT.

Problem~\eqref{eq:PRoptim} resembles the phase synchronization problem~\cite{singer2011angular,bandeira2014tightness,boumal2016nonconvex}. In~\cite{waldspurger2015phase}, the authors pursue a convex relaxation of~\eqref{eq:PRoptim} named \emph{phasecut}. Here, following~\cite{boumal2016nonconvex}, we use the Manopt toolbox to run local optimization of~\eqref{eq:PRoptim} over the manifold of phases~\cite{manopt}. {In its simplest form, the algorithm follows the gradient's component  which is consistent with the feasible set of solutions (see details below).}
To initialize the local optimization algorithm, we set $\mathbf{U}_0$ to be the phases of $\STFT\left(\mathbf{x}_0\right)$, where $\mathbf{x}_0$ is the initialization used by Algorithm~\ref{alg:non_convex}.
This approach is summarized in Algorithm~\ref{alg:PR}.

For completeness, we provide a brief overview of step~\ref{item:algo_PR_optim} of Algorithm~\ref{alg:PR}, that is, optimization of the phases. We restrict attention to a simple Riemannian optimization algorithm, namely, the gradient ascent algorithm. See~\cite{genrtr} for details about the more sophisticated Riemannian trust-region method (RTR), which we use in practice.

The variable $\mathbf{U}$ lives on a smooth manifold, namely, the set of phases
\begin{align*}
\mathcal{M} & = \{ \mathbf{U} \in \mathbb{C}^{\left\lceil \frac{N}{L}\right\rceil\times N} : |\mathbf{U}[m,k]| = 1 \textrm{ for all } m,k\},
\end{align*}
which is a Cartesian product of unit circles in the complex plane (a torus). This smooth nonlinear space can be linearized about every point $\mathbf{U}$  by differentiating the constraints. This yields a linear subspace known as the tangent space to $\mathcal{M}$ at~$\mathbf{U}$:
\begin{align*}
\mathrm{T}_\mathbf{U} \mathcal{M} & = \{ \dot{\mathbf{U}} \in \mathbb{C}^{\left\lceil \frac{N}{L}\right\rceil\times N} : \Re\{ \overline{\mathbf{U}} \odot \dot {\mathbf{U}} \} = 0 \}.
\end{align*}
Each tangent space    of $\mathcal{M}$ can be endowed with the  inner product~\eqref{eq:frobinner} (simply by restricting it to each particular subspace), which turns $\mathcal{M}$ into a Riemannian submanifold of $\mathbb{C}^{\left\lceil \frac{N}{L}\right\rceil \times N}$. This makes it particularly easy to compute the gradient of the objective function $f \colon \mathcal{M} \to \mathbb{R}$,
\begin{align} \label{eq:f}
f(\mathbf{U}) & = \left\langle \mathbf{Z}^{1/2} \odot \mathbf{U}, \STFT \circ \STFTdagger (\mathbf{Z}^{1/2} \odot \mathbf{U}) \right\rangle.
\end{align}

Indeed, following~\cite[eq.~(3.37)]{AMS08}, the gradient of $f$ at $\mathbf{U}$ \emph{restricted to $\mathcal{M}$}---known as the Riemannian gradient $\operatorname{grad} f(\mathbf{U})$---is the orthogonal projection of the classical (unconstrained) gradient of $f$, denoted by $\nabla f(\mathbf{U})$, to the tangent space $\mathrm{T}_\mathbf{U} \mathcal{M}$:
\begin{align}
\nabla f(\mathbf{U}) & = 2\mathbf{Z}^{1/2} \odot \left(\STFT \circ \STFTdagger (\mathbf{Z}^{1/2} \odot \mathbf{U})\right), \nonumber\\
\operatorname{grad} f(\mathbf{U}) & = \operatorname{Proj}_\mathbf{U}(\nabla f(\mathbf{U})).
\label{eq:grads}
\end{align}
The orthogonal projector $\operatorname{Proj}_\mathbf{U} \colon \mathbb{C}^{ \left\lceil \frac{N}{L}\right\rceil \times N} \to \mathrm{T}_\mathbf{U} \mathcal{M}$ is given by
\begin{align*}
\operatorname{Proj}_\mathbf{U}(\mathbf{V}) & = \mathbf{V} - \Re\{ \overline{\mathbf{U}} \odot \mathbf{V} \} \odot \mathbf{U}.
\end{align*}
That is, it subtracts from each entry $\mathbf{V}[m,k]$ its component aligned with $\mathbf{U}[m,k]$.
Explicitly, the Riemannian gradient is then given by
\begin{equation*}
\operatorname{grad} f(\mathbf{U}) = \nabla f(\mathbf{U}) - \Re\{ \overline{\mathbf{U}} \odot \nabla f(\mathbf{U}) \} \odot \mathbf{U}.
\end{equation*}

Now that we are equipped with a notion of gradient on the manifold, the only missing ingredient to implement a gradient ascent optimization algorithm is a means of moving away from a point (a current iterate) along a chosen tangent direction (here, the gradient vector), while remaining on the manifold $\mathcal{M}$. The standard tool to achieve this is known as a \emph{retraction}~\cite[\S4.1]{AMS08}. An obvious retraction for $\mathcal{M}$ is
\begin{align*}
\operatorname{Retr}_\mathbf{U}(\dot{\mathbf{U}}) & = \operatorname{phase}(\mathbf{U} + \dot{\mathbf{U}}).
\end{align*}
Indeed, for $\mathbf{U} \in \mathcal{M}$ and $\dot{\mathbf{U}} \in \mathrm{T}_\mathbf{U}\mathcal{M}$, the result of this operation is always on $\mathcal{M}$ and locally (that is, for small $\dot{\mathbf{U}}$) the change is along the prescribed tangent direction $\dot{\mathbf{U}}$. 

The gradient ascent algorithm takes the form
\begin{align} \label{eq:retraction}
\mathbf{U}_{k+1} & = \operatorname{Retr}_{\mathbf{U}_k}(\eta_k \operatorname{grad} f(\mathbf{U}_k)),
\end{align}
where $\eta_k > 0$ is an appropriately chosen step size (typically using a form of line-search~\cite[\S4]{AMS08}) and $\mathbf{U}_0 \in \mathcal{M}$ is a given initial guess. Owing to $\mathcal{M}$ being a compact submanifold of $\mathbb{C}^{\left\lceil \frac{N}{L}\right\rceil \times N}$ and to $f$ being smooth, both Riemannian gradient ascent (with appropriate line-search) and RTR are guaranteed to converge to points which satisfy first-order necessary optimality conditions, that is, $\| \operatorname{grad} f(\mathbf{U}) \| = 0$ (and even second-order conditions for RTR) regardless of initialization, with known worst-case bounds on iteration counts~\cite{boumal2016globalrates}. Explicitly, at a critical point $ \mathbf{U}$ the algorithm satisfies:  
\begin{equation} \label{eq:ncpc_stop}
 \mathbf{U} = \operatorname{phase}\left(\STFT \circ \STFTdagger (\mathbf{Z}^{1/2} \odot \mathbf{U})\right).
\end{equation}
As will be shown next, this is also the stagnation point of Fienup's algorithm.
 This approach is summarized in Algorithm~\ref{alg:RGD}. 

We stress that this approach is different from a projected gradient method. Indeed, in a projected gradient method, one would alternate between following the classical gradient $\nabla f(\mathbf{U})$ and projecting to $\mathcal{M}$ with the $\operatorname{phase}$ operator. That is, each iteration resembles~\eqref{eq:retraction} with  $\nabla f$ instead of $\operatorname{grad} f$.
In contrast, the Riemannian gradient method follows the tangent part of the gradient, $\operatorname{grad} f$~\eqref{eq:grads} and then projects onto $\mathcal{M}$. One advantage is that, close to convergence, the Riemannian gradient has small norm (as expected), whereas the classical gradient may still be large.

\begin{algorithm}
	\textbf{Input:} The measurements $\mathbf{Z} \approx |\STFT(\mathbf{x})|^2$ as given in (\ref{eq:meas0}).\\
	\textbf{Output:} $\hat{\mathbf{x}}$: estimation of \textbf{$\mathbf{x}$}.
	\begin{enumerate}
		\item Compute the initialization $\mathbf{x}_0$ with Algorithm~\ref{alg:non_convex} to obtain $\mathbf{U}_0 = \operatorname{phase}(\STFT(\mathbf{x}_0))$.
		\item Using $\mathbf{U}_0$ as initialization, use a local optimization algorithm to try to compute a solution $\mathbf{\hat U}$ to
		\begin{align*}
		\max_{\mathbf{U} \in \mathbb{C}^{ \left\lceil \frac{N}{L}\right\rceil \times N}} & \left\langle \mathbf{Z}^{1/2} \odot \mathbf{U}, \STFT \circ \STFTdagger (\mathbf{Z}^{1/2} \odot \mathbf{U}) \right\rangle \\ \textrm{ subject to } & \quad |\mathbf{U}[m,k]| = 1,\thinspace \forall m, k.
		\end{align*}
		See Algorithm~\ref{alg:RGD} for a simple Riemannian gradient method; see~\cite{genrtr,manopt} for Riemannian trust regions. \label{item:algo_PR_optim}
		\item Return $\mathbf{\hat x} = \STFTdagger(\mathbf{Z}^{1/2} \odot \hat{\mathbf{U}})$.
	\end{enumerate}
	\protect\caption{\label{alg:PR} Non-Convex PhaseCut (NCPC)}
\end{algorithm}

\begin{algorithm}
	\textbf{Input:} The measurements $\mathbf{Z} \approx |\STFT(\mathbf{x})|^2$ as given in (\ref{eq:meas0}), initial guess $\mathbf{U}_0 \in \mathcal{M}$ and tolerance $\varepsilon > 0$.\\
	\textbf{Output:} $\hat{\mathbf{U}} \in \mathcal{M}$ satisfying $\|\operatorname{grad} f(\hat{\mathbf{U}})\|_{\mathrm{F}} \leq \varepsilon$.
	\begin{enumerate}
		\item[] For $k = 0, 1, \ldots$
		\begin{enumerate}
			\item Compute: $$ \operatorname{grad} f(\mathbf{U}_k) = \nabla f(\mathbf{U}_k) - \Re\{ \overline{\mathbf{U}_k} \odot \nabla f(\mathbf{U}_k) \} \odot \mathbf{U}_k,
			$$  where $$ \nabla f(\mathbf{U}_k) = 2 \mathbf{Z}^{1/2} \odot \left(\STFT \circ \STFTdagger (\mathbf{Z}^{1/2} \odot \mathbf{U}_k)\right).$$
			\item If $\|\operatorname{grad} f(\mathbf{U}_k)\|_{\mathrm{F}} \leq \varepsilon$, return $\hat{\mathbf{U}} = \mathbf{U}_k.$
			\item Compute a step size $\eta_k$ with a classical line-search algorithm, e.g.,~\cite[\S4]{AMS08}.
			\item Set $\mathbf{U}_{k+1} = \operatorname{phase}(\mathbf{U}_k + \eta_k \operatorname{grad} f(\mathbf{U}_k))$.
		\end{enumerate}
	\end{enumerate}
	\protect\caption{\label{alg:RGD} Riemannian gradient method for NCPC}
\end{algorithm}


\subsubsection{Relation to Fienup's Algorithm} \label{sec:relation_to_Fienup}

Our method can be compared with the classical Fienup algorithm for the STFT case, also called Griffin--Lim algorithm~\cite{griffin1984signal}, as follows. One approach to optimize~\eqref{eq:PRoptim}, instead of the Riemannian gradient iterations that we describe in Algorithm~\ref{alg:RGD}, is an iterative technique called projected power method (PPM), or generalized power method~\cite{journee2010generalized,boumal2016nonconvex}. This algorithm iterates as the power method, with the difference that, at each iteration, it keeps only the phases of the current iterate. Specifically, the $k$th iteration is of the form 
	\begin{equation} \label{eq:ppm}
	\mathbf{U}_{k}=\operatorname{phase}\left(\STFT\circ\STFTdagger\left(\mathbf{U}_{k-1}\odot\mathbf{Z}^{1/2}\right)\right).
	\end{equation}	
	Similarly to NCPC, the algorithm stops when~\eqref{eq:ncpc_stop} is satisfied.
	On the other hand, each iteration of Fienup's algorithm takes on the form 
	\begin{equation}\label{eq:fienup_dagger}
	\mathbf{{x}}_{k} = \STFTdagger\left(\operatorname{phase}\left(\STFT(\mathbf{{x}}_{k-1})\right)\odot\mathbf{Z}^{1/2} \right).
	\end{equation}
	Applying the operator $\operatorname{phase}\circ \STFT$ on the iterations  of~\eqref{eq:fienup_dagger} shows that it is equivalent to PPM through the mapping $\mathbf{U}_{k} =\operatorname{phase}\circ\STFT(\mathbf{x}_{k})$. 
	In this sense, one can understand Fienup's algorithm  as a particular iterative method to solve the optimization problem~\eqref{eq:PRoptim}.

	According to~\cite[Lemma 15]{boumal2016nonconvex}, all fixed points of~\eqref{eq:ppm}--and hence of~\eqref{eq:fienup_dagger}--map to critical points of the optimization problem~\eqref{eq:PRoptim}, that is, they map to points $\mathbf{U}_k$ where the Riemannian gradient  is zero. These are only the first-order necessary optimality conditions. Numerical experiments (not displayed here) show that some of the stable fixed points of~\eqref{eq:fienup_dagger} map to critical points which do not satisfy the second-order necessary optimality conditions (their Riemannian Hessian admits a positive eigenvalue) and  are therefore suboptimal. In contrast, such points would be unstable fixed points for any reasonable Riemannian optimization algorithm  as confirmed in the same experiments. This distinction at least partially explains why the empirical performance of the NCPC algorithm is superior to that of  Fienup's algorithm, as demonstrated in Section~\ref{sec:Numerical-Results}.


\subsection{Initialization}


\subsubsection{Initialization for $L=1$}

 Since the phase retrieval problem is inherently non-convex, it is not clear whether the proposed refinement algorithms 
will converge to a global minimum  from an arbitrary initialization. When $L=1$, we propose initializing the iterations by using Algorithm \ref{alg:linearLS}.
As explained in Section \ref{sec:uniqueness}, for $W\geq \left\lceil \frac{N+1}{2}\right\rceil$ the algorithm returns $\mathbf{x}$ exactly. 
 However, when $W< \left\lceil\frac{N+1}{2}\right\rceil$, $\mathbf{G}_\ell=0$ for $\ell=W,\dots,N-W$ so that the output is not necessarily $\mathbf{x}$.  Nevertheless, in Section \ref{sec:theory} we provide theoretical guarantees establishing that under  appropriate conditions, this initialization results in a good approximation.

In practical applications,   a variety of approaches are  used to initialize the refinement techniques. While the specific initialization method is application-dependent, these approaches can be broadly classified into two categories. The first is based on the structure of the expected signal. For instance, in some applications it is common to use a Gaussian pulse with random phases  as an initial point~\cite{delong1994frequency}. This, however, may lead to a phenomenon called \emph{model bias} in which the estimate tends to capture characteristics of the model rather than the true signal.
 An alternative strategy, also used by commercial software,  is based on random initialization. This is very different from our initialization  which exploits the acquired data.


\subsubsection{Initialization for $L>1$}

Until now we focused  on maximal overlap between adjacent windows, namely, $L=1$. When $L>1$,
(\ref{eq:Gl}) results in an underdetermined system of equations since
$\mathbf{y}_{\ell}\in\mathbb{R}^{\left\lceil \frac{N}{L}\right\rceil}$, $\mathbf{G}_{\ell}\in\mathbb{R}^{\left\lceil \frac{N}{L}\right\rceil\times N}$
and $\mathbf{x}_{\ell}\in\mathbb{R}^{N}$.  In this case, the LS solution $\mathbf{G}_{\ell}^{\dagger}\mathbf{y}_{\ell}$ is the vector with minimal $\ell_2$ norm among the set of feasible solutions. This approximation is quite poor in general.

We notice that the measurements $\mathbf{y}_{\ell}$ are a downsampled version by a factor  $L$ of the case of maximal overlap $(L=1)$. 
Therefore, we suggest  upsampling $\mathbf{y}_{\ell}$ to approximate the  maximal overlap setting based on the averaging nature of the window $\mathbf{g}$. 
In order to motivate our approach, we start by considering an ideal
situation. Suppose that for some $\ell$, the DFT of the first column of $\mathbf{G}_\ell$, denoted by $\mathbf{\hat{g}}_\ell$,  
is an ideal low-pass with integer bandwidth $N/L_{BW}$. Namely,
\[
\mathbf{\hat{g}}_\ell[k]=\begin{cases}
1,\quad & k=0,\dots  N/L_{BW} -1,\\
0,\quad & \mbox{otherwise}.
\end{cases}
\]
The following lemma states that in this case,
 no information is lost
by choosing  $L=L_{BW}$ compared to taking maximal overlap $L=1$. Moreover, it suggests to upsample the measurement vector by expansion and low-pass interpolation.  Our technique   resembles standard upsampling arguments in digital signal
processing (DSP) (see for instance Section 4.6 of \cite{oppenheim2010discrete}).

\begin{lem}
\label{claim_interpolation} Let $\mathbf{\tilde{g}}:=\left\{\mathbf{g}[(-n)\bmod N ]\right\}_{n=0}^{N-1} $. Suppose that  $\mathbf{\tilde{g}}\in\mathbb{R}^{N}$ 
is an ideal low-pass with integer bandwidth  $ {N}/{L}$
and $\mathbf{y}=\mathbf{g}\ast \mathbf{x}$ for some $\mathbf{x}\in\mathbb{C}^N$ (or equivalently, $\mathbf{y}=\mathbf{G} \mathbf{x}$, where $\mathbf{G}$ is a circualnt matrix whose first column is $\mathbf{\tilde{g}}$).  Let  $\mathbf{y}_L\in\mathbb{C}^{ \frac{N}{L}}$ be its $L$-downsampled version, i.e.,
\[ 
\mathbf{y}_L[n]=\mathbf{y}[nL],\quad n=0,\dots, N/L -1.
\]
Then, $\mathbf{y}=\left(\mathbf{F}_{p}^{*}\mathbf{F}_{p}\right)\tilde{\mathbf{y}}_{L}$,
where 
\begin{equation}
{\mathbf{\tilde{y}}}_{L}[n]=\begin{cases}
\mathbf{y}_{L}[m], & n=mL,\\
0, & \mbox{otherwise,}\label{eq:expansion}
\end{cases}
\end{equation}
and $\mathbf{F}_{p}$ is a partial Fourier matrix consisting of the
first $ N/L  $ rows of the DFT matrix $\mathbf{F}$. \end{lem}
\begin{IEEEproof}
See Appendix \ref{sec:proof_of_interpolation}. 
\end{IEEEproof}

While Lemma \ref{claim_interpolation} shows that no information is lost using an ideal low-pass window with integer bandwidth $ N/L $, in practice we do not use these windows. Instead, 
we approximate the low-pass interpolation of $\mathbf{F}_{p}^{*}\mathbf{F}_{p}$
as suggested in Lemma \ref{claim_interpolation} by a simple smooth interpolation.
This leads to  better numerical results and reduces the computational complexity.
In Section \ref{sec:Numerical-Results} we show simulations with
both linear and cubic interpolations.

Following  the upsampling stage, the algorithm proceeds as for $L=1$ by extracting the principal eigenvector (with the appropriate normalization) of an approximation matrix.
This initialization  is summarized in Algorithm \ref{alg:linearLS_L}.

\begin{algorithm}
\textbf{Input:} The measurements $\mathbf{Z}[m,k]$ as given in (\ref{eq:meas0})
and a smooth interpolation filter $\mathbf{h}_{L}\in\mathbb{R}^N$ that approximates a low-pass filter with bandwidth $\left\lceil N/L \right\rceil $.

\textbf{Output:} $\mathbf{x}_{0}$: Estimation of \textbf{$\mathbf{x}$}. 
\begin{enumerate}
\item Compute $\mathbf{Y}\left[m,\ell\right]$, the 1D DFT with respect to the second variable of $\mathbf{Z}[m,k]$
as given in (\ref{eq:meas}).
\item Upsampling: For each $\ell\in[-(W-1),\dots,(W-1)]$:
\begin{enumerate}
\item Let $\mathbf{{y}}_\ell[m]:=\left\{\mathbf{Y}\left[m,\ell\right]\right\}_{m=0}^{\left\lceil \frac{N}{L}\right\rceil-1}$. 
\item Expansion:
\[
{\mathbf{\tilde{y}}}_{\ell}[n]:=\begin{cases}
\mathbf{{y}}_\ell[m], & \quad n=mL,\\
0, & \quad\mbox{otherwise}.
\end{cases}
\]

\item Interpolation:
\[
\mathbf{\bar{y}}_{\ell}=\mathbf{\tilde{y}}_{\ell}\ast\mathbf{h}_{L}.
\]

\end{enumerate}

\item Construct a matrix $\mathbf{X}_{0}$ such that 
\[
\diag\left(\mathbf{X}_{0},\ell\right)=\begin{cases}
\mathbf{G}_{\ell}^{\dagger}\mathbf{\bar{y}}_\ell, & \ell=-\left(W-1\right),\cdots,\left(W-1\right),\\
0, & \mbox{otherwise,}
\end{cases}
\]
where $\mathbf{G}_{\ell}\in\mathbb{R}^{N\times N}$ are defined as  in (\ref{eq:Gl}) for $L=1$. 
\item Let $\mathbf{x}_{p}$ be the principal (unit-norm) eigenvector of
$\mathbf{X}_{0}$. Then, 
\[
\mathbf{x}_{0}=\sqrt{\sum_{n\in P}\left(\mathbf{G}_{0}^{\dagger}\mathbf{y}_{0}\right)[n]}\mathbf{x}_{p},
\]
where $P:=\left\{ n\thinspace:\thinspace\left(\mathbf{G}_{0}^{\dagger}\mathbf{y}_{0}\right)[n]>0\right\} $. 
\end{enumerate}
\protect\caption{\label{alg:linearLS_L}Least-squares initialization for $L>1$}
\end{algorithm}


\section{\label{sec:Numerical-Results}Numerical Results}

This section is devoted to  numerical experiments examining
the proposed non-convex algorithms. In all experiments, the underlying signal was drawn
from $\mathbf{x}\sim\mathcal{N}\left(0,\mathbf{I}\right)$, where
$\mathbf{I}$ is the identity matrix. { The measurements $\mathbf{Z}$~\eqref{eq:meas0} were  contaminated with either   i.i.d.\ additive Gaussian noise  or Poisson noise.}
 The recovery error is computed by $\frac{d\left(\mathbf{x},\hat{\mathbf{x}}\right)}{\left\Vert \mathbf{x}\right\Vert _{2}}$,
where $\hat{\mathbf{x}}$ is the estimated signal and the distance function $d\left(\cdot,\cdot\right)$ is defined in Definition \ref{def:dis}.
 We optimize both the empirical risk loss function \eqref{eq:LS_Z} and the non-convex phasecut (NCPC) objective function by a trust-region algorithm using the Manopt toolbox \cite{manopt}.

\begin{figure*}
\begin{minipage}[t]{1.1\columnwidth}%
\subfloat[Initialization with linear interpolation]{\includegraphics[scale=0.6]{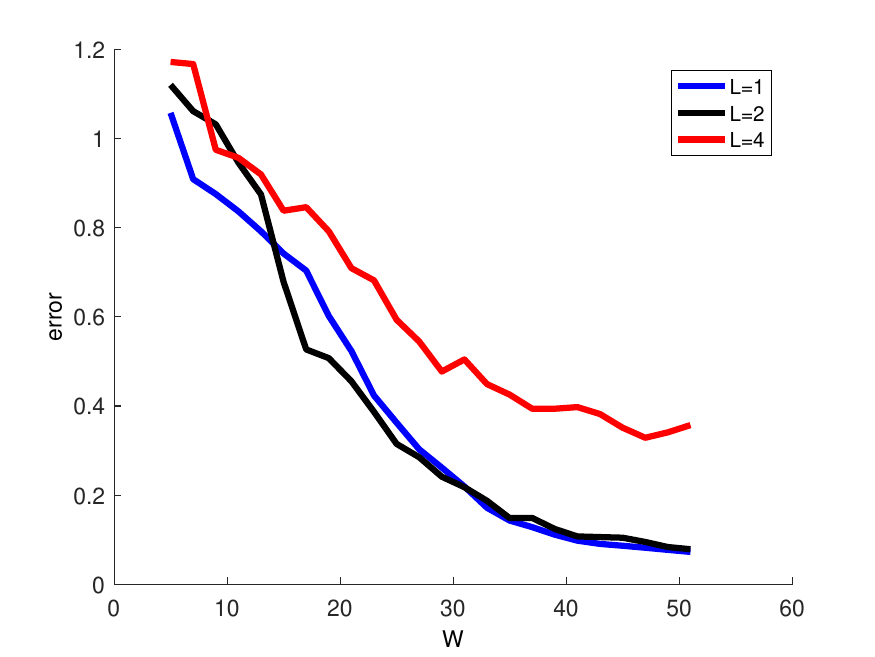}

}%
\end{minipage}%
\begin{minipage}[t]{1.1\columnwidth}%
\subfloat[Initialization with cubic interpolation]{\includegraphics[scale=0.6]{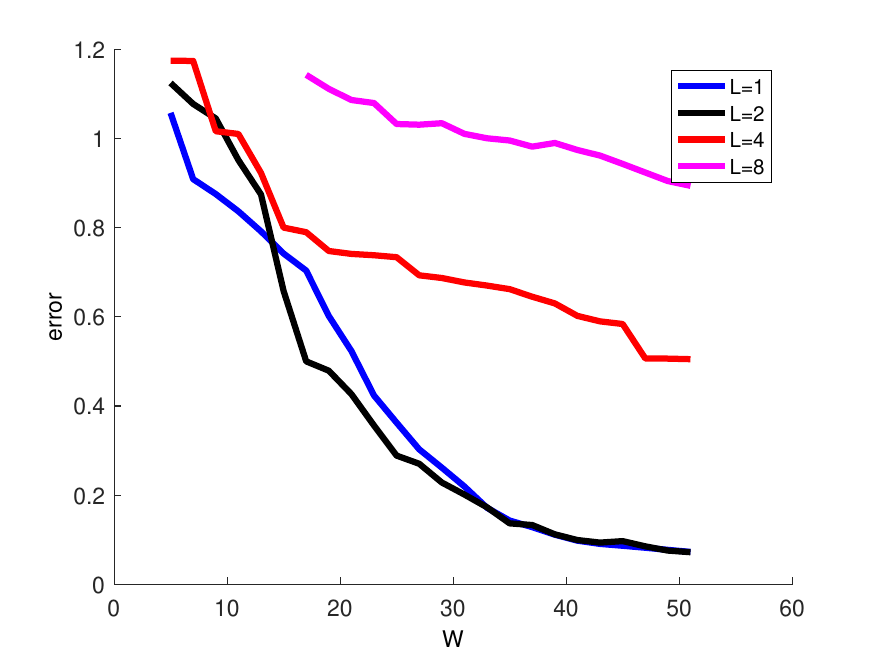}

}%
\end{minipage}

\protect\caption{\label{fig:initial_comp}Average error (over 50 experiments) of
the initialization of Algorithm \ref{alg:linearLS_L} as a function of
$W$ and $L$. The experiments were conducted on a signal of length
$N=101$ with a Gaussian window $e^{-\frac{n^{2}}{2\sigma^{2}}}$
and linear or cubic interpolation. The window length was set to be
$W=3\sigma$. }
\end{figure*}

The first experiment examines the estimation quality of the initialization method 
described in Algorithm~\ref{alg:linearLS_L}. Figure~\ref{fig:initial_comp}
presents the initialization error as a function of the window's length.
We considered a Gaussian window defined by $\mathbf{g}[n]=e^{\frac{-n^{2}}{2\sigma^{2}}}$
and cubic and linear interpolations. For $n>3\sigma$, we set the entries of the window to be zero
so that $W=3\sigma$. The results demonstrate the effectiveness of the 
smooth interpolation technique. For low values of $L$, it seems that
the two interpolations achieve similar performance. For larger $L$,
namely, fewer measurements, cubic interpolation outperforms  linear interpolation. In the following experiments we use  cubic interpolation.

\begin{figure}
\begin{centering}
\includegraphics[scale=0.6]{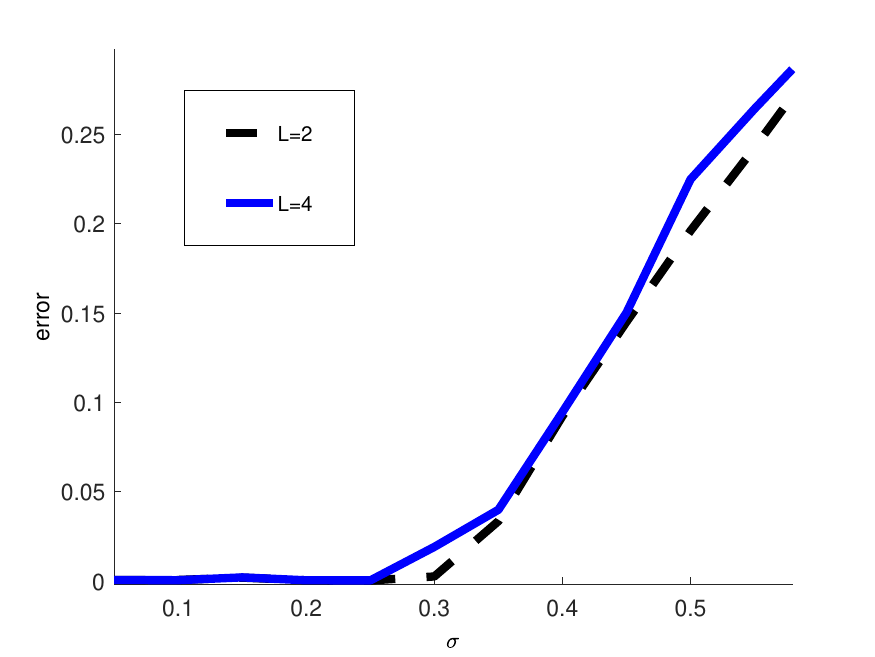} 
\par\end{centering}
\protect\caption{\label{fig:basin_of_attraction} Average recovery error (over 100 experiments)
	of minimizing the ER loss function~\eqref{eq:LS_Z} or~\eqref{eq:LS} for signals of length  $N=43$ and a rectangular window of length $W=11$. The initialization   was set as $\mathbf{x}_{0}=\mathbf{x}+\mathbf{z}$,
where $\mathbf{x}$ is the underlying signal and the perturbation vector $\mathbf{z}$ takes the values of $\pm\sigma$ for some $\sigma>0$ where the sign is drawn randomly.}
\end{figure}

The next experiment aims to estimate the 
\emph{basin of attraction} of the loss function~\eqref{eq:LS_Z} or~\eqref{eq:LS}. That is to say, the
area in which a local optimization method
 will converge to a global minimum. 
To do that, we set the initialization vector  to be $\mathbf{x}_{0}=\mathbf{x}+\mathbf{z}$, where $\mathbf{x}\sim\mathcal{N}\left(0,\mathbf{I}\right)$
is the underlying signal. The perturbation vector  $\mathbf{z}$ takes on the values  $\pm \sigma$ (with random signs) for some $\sigma>0$  so that $d(\mathbf{x}_{0},\mathbf{x})\leq \sqrt{N}\sigma$.
Then, we applied the trust-region algorithm and checked whether the algorithm
converges to $\mathbf{x}$. As can be seen in Figure \ref{fig:basin_of_attraction},
the algorithm converges to the global minimum as long as $\sigma\leq 0.3$ for $L=1,2$ (the case of $L=1$ is not presented in the figure) and $\sigma\leq 0.25$ for $L=4$. These experimental results indicate that the actual basin of attraction is larger than our theoretical estimation  in Section \ref{sec:theory} and  Theorem \ref{lem:converges}.

\begin{figure*}
	\begin{minipage}[t]{1\columnwidth}%
		\subfloat[Initialization with $W=7$ and $L=1$]{\includegraphics[scale=0.4]{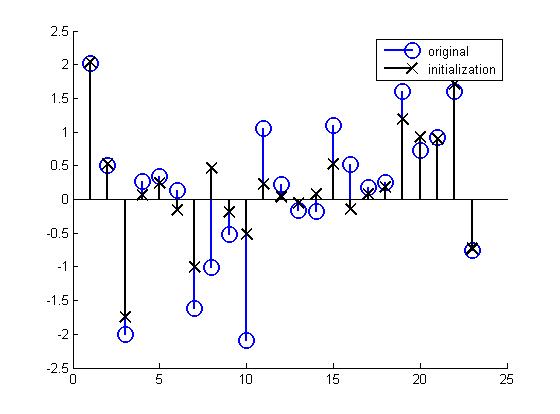}			
		}%
	\end{minipage}%
	\begin{minipage}[t]{1\columnwidth}%
		\subfloat[Initialization with $W=11$ and $L=3$]{\includegraphics[scale=0.4]{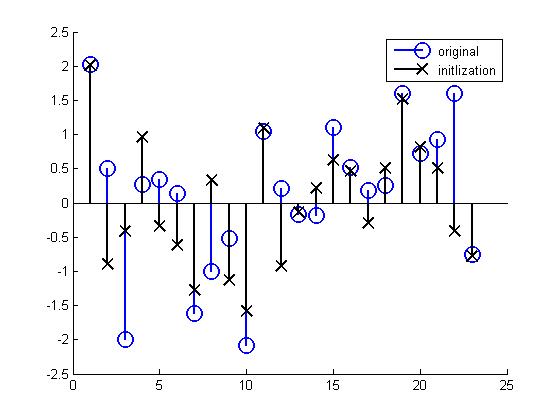}			
		}%
	\end{minipage}
	
	\begin{minipage}[t]{1\columnwidth}%
		\subfloat[Recovery with $W=7$ and $L=1$]{\includegraphics[scale=0.4]{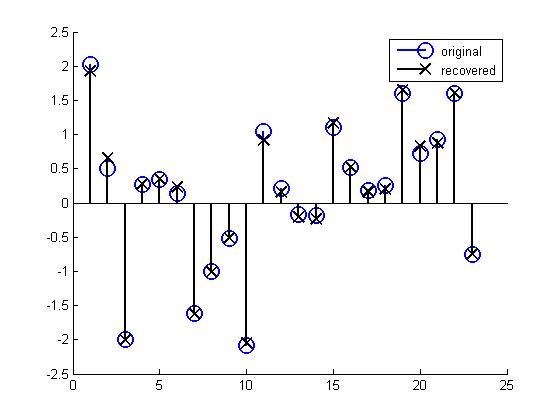}			
		}%
	\end{minipage}%
	\begin{minipage}[t]{1\columnwidth}%
		\subfloat[Recovery  with $W=11$ and $L=3$]{\includegraphics[scale=0.4]{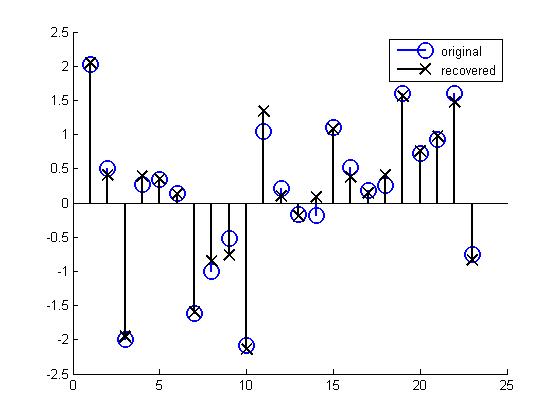}
			
		}%
	\end{minipage}
	\protect\caption{\label{fig:Example}Recovery of a signal of length $N=23$ with a
		rectangular window in a noisy environment of SNR$=20$ dB. We used a trust-region algorithm to minimize the ER loss function \eqref{eq:LS_Z}.
		 The experiments were conducted with
		$W=7$ and $L=1$ and $W=11$ and $L=3$ in the left and the right
		columns, respectively.}
\end{figure*}

Figure \ref{fig:Example} shows a representative example of the
performance of Algorithm \ref{alg:non_convex} where we minimized the empirical risk loss function~\eqref{eq:LS_Z} or~\eqref{eq:LS}. The experiment was conducted
on a signal of length $N=23$ with a rectangular window in a noisy environment
of SNR$=20$ dB. 

\begin{figure}
	\begin{centering}
		\includegraphics[scale=0.6]{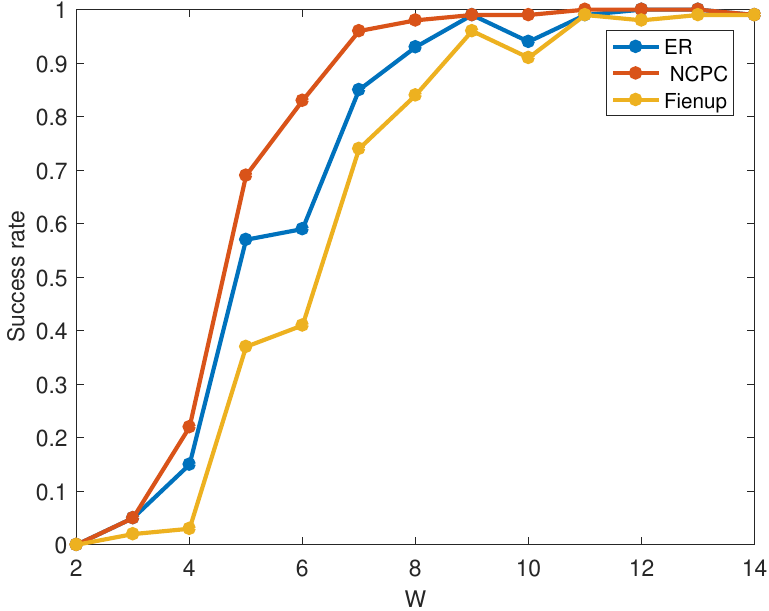} 
		\par\end{centering}	
	\protect\caption{\label{fig:success_rate} Success rate as a function of $W$ over 100 experiments conducted with $N=31$, $L=2$ and a rectangular window. We compared three algorithms: minimizing the ER loss function \eqref{eq:LS_Z}, NCPC and Fienup. A success was declared for recovery error  less than $10^{-3}$.}
\end{figure}

\begin{figure*}
	\centering
	\begin{minipage}[t]{1.1\columnwidth}%
		\subfloat[Recovery error with Gaussian i.i.d.\ noise as a function of the SNR with $W=15$.]{\includegraphics[scale=0.6]{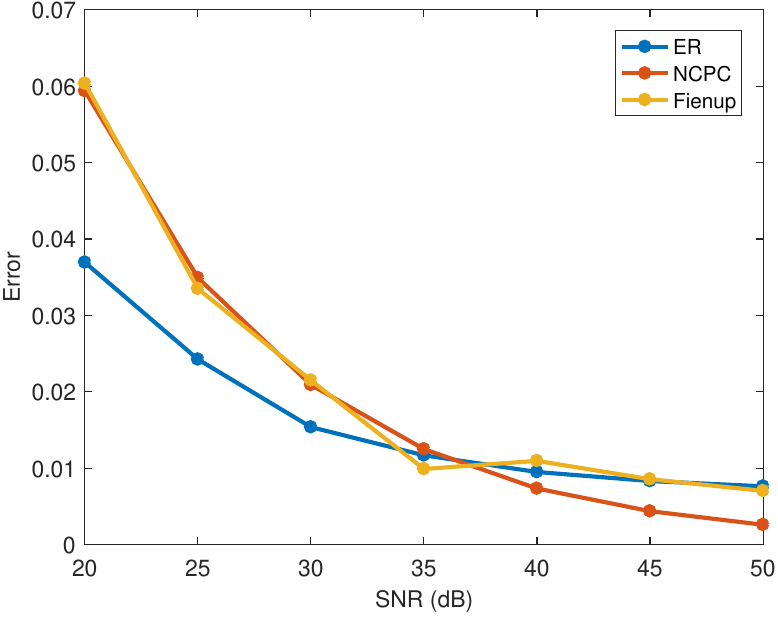}			
		}%
	\end{minipage}%
	\begin{minipage}[t]{1.1\columnwidth}%
		\subfloat[Recovery error with Poisson noise as a function of the window's length $W$.]{\includegraphics[scale=0.6]{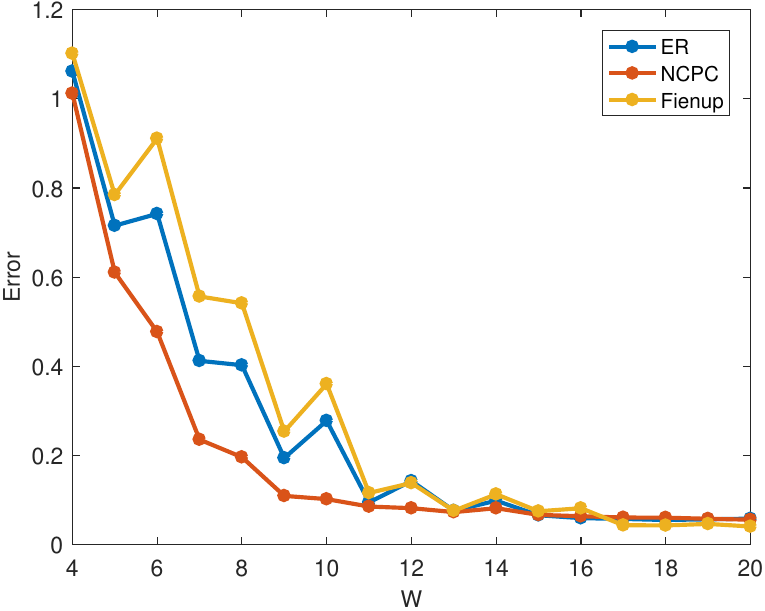}			
		}%
	\end{minipage}	
	\protect\caption{\label{fig:stability} Comparison of the  average recovery error (over 100 experiments)
		of three algorithms: minimizing the ER loss function~\eqref{eq:LS_Z}, NCPC and Fienup \cite{griffin1984signal}. The experiments
		were conducted on signals of length $N=53$ with a rectangular window and $L=2$. }
\end{figure*}

\begin{figure*}
	\centering
	\begin{minipage}[t]{1.1\columnwidth}%
		\subfloat[Recovery error with Gaussian i.i.d.\ noise as a function of the SNR with~$W=15$.]{\includegraphics[scale=0.16]{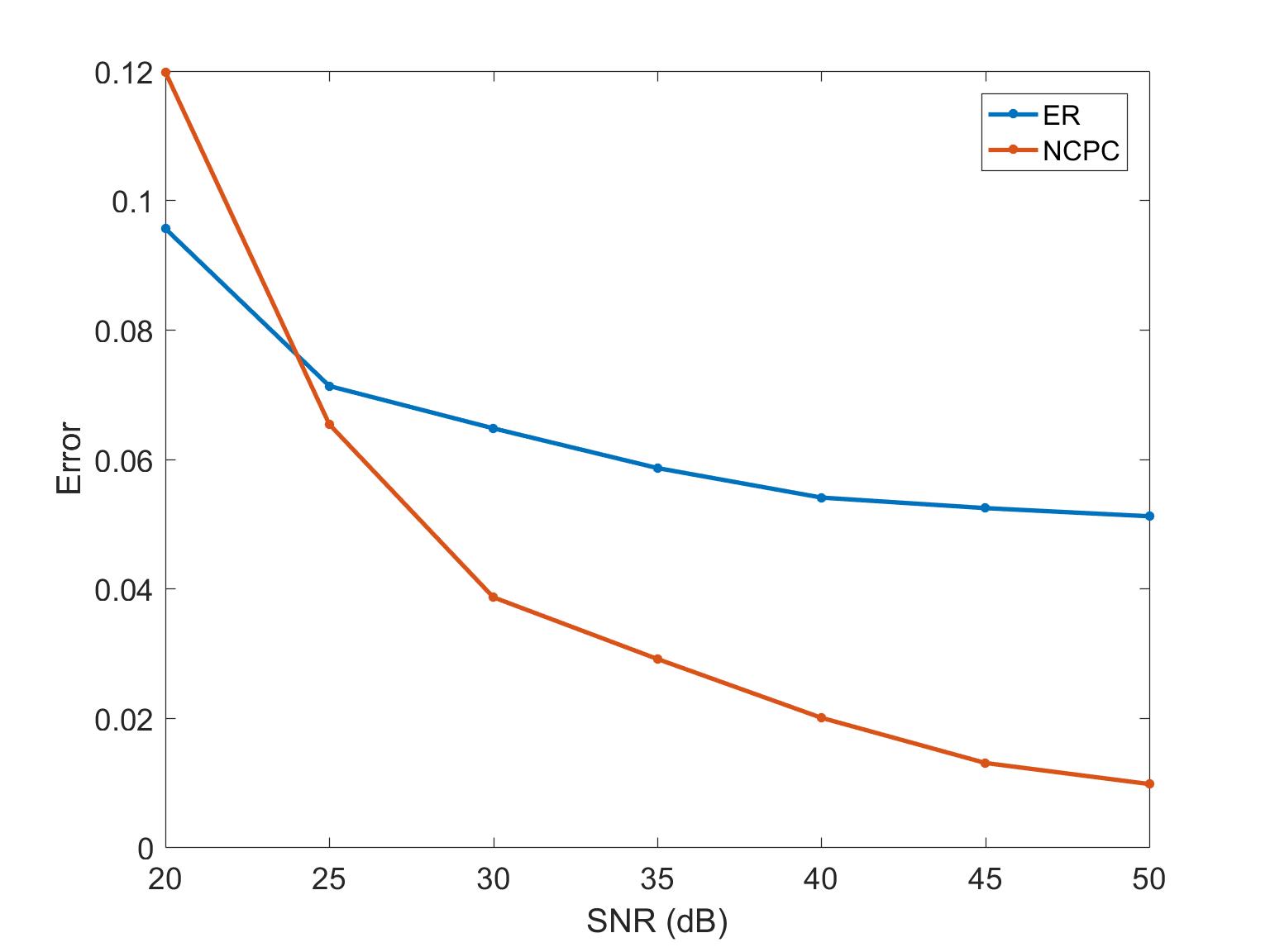}			
		}%
	\end{minipage}%
	\begin{minipage}[t]{1.1\columnwidth}%
		\subfloat[Recovery error with Poisson noise as a function of the window's length~$W$.]{\includegraphics[scale=0.16]{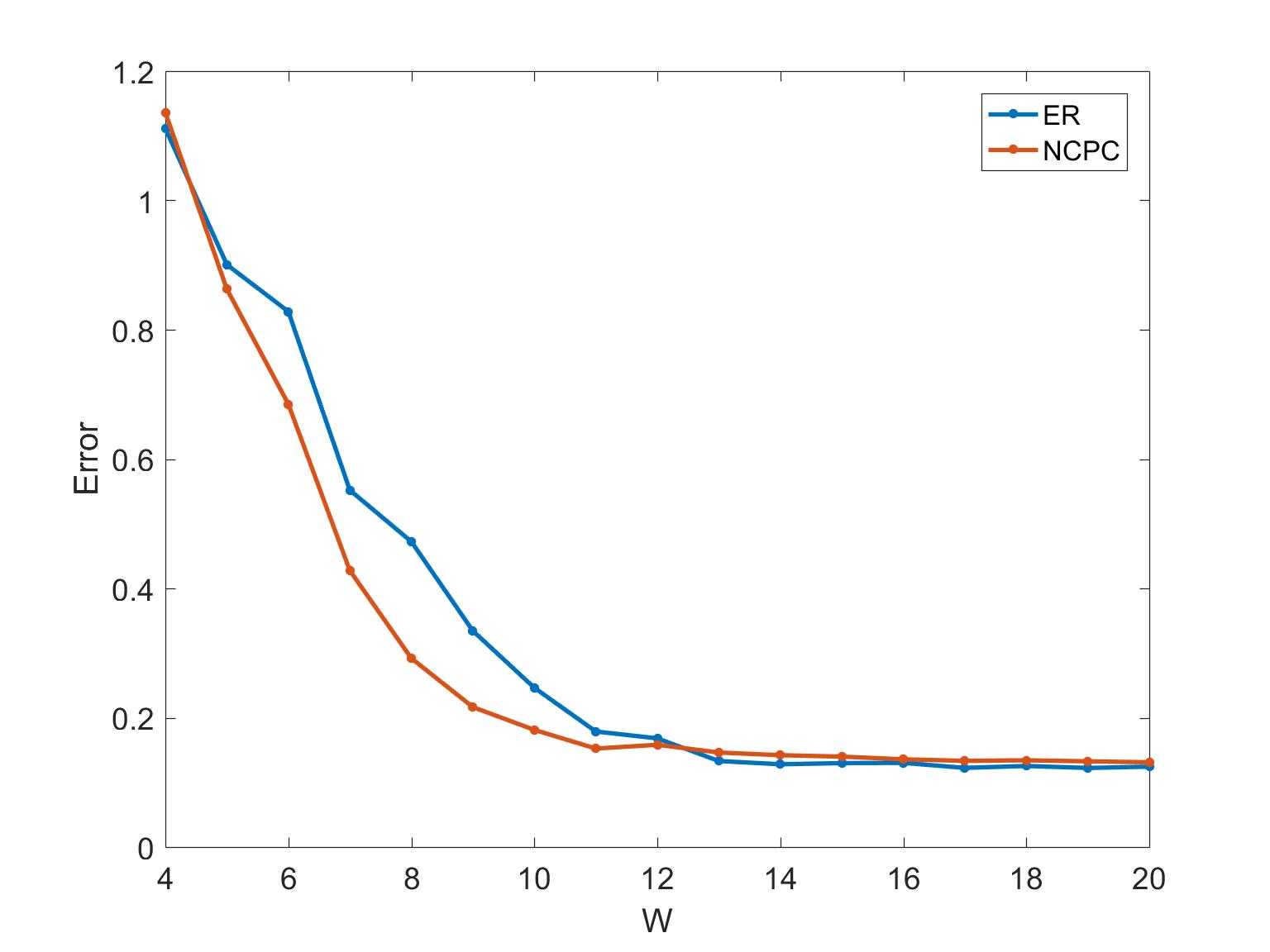}			
		}%
	\end{minipage}	
	\protect\caption{\label{fig:stability_LP} Average recovery error (over 100 experiments with signals of length $N=53$, a rectangular window and $L=2$)
		of minimizing the ER loss function \eqref{eq:LS_Z} and NCPC with low-passed data. Particularly, we used the measured data $\mathbf{Z}[m,k]$ for all $m$ and $k=-K_{\max},\dots,K_{\max}$ with $K_{\max}=18$.}
\end{figure*}

Figure \ref{fig:success_rate} presents the success rate of the algorithms as a function of the window's length in a noise-free environment. As can be seen,  NCPC achieves  the highest success rate, implying that it requires less redundancy in the data. 
  Figure \ref{fig:stability} presents the recovery error for different noise models. Figure \ref{fig:stability}a shows the error when the measurements are contaminated with normal noise as a function of the  SNR level. The proposed algorithms are compared with Fienup's method~\cite{griffin1984signal} that iterates according to~\eqref{eq:fienup_dagger}. In the low SNR regime, minimizing the ER loss function seems to be better. Figure \ref{fig:stability}b shows the error with Poisson noise as a function of $W$. For short windows, NCPC works best. The performance for longer windows is comparable for all  algorithms. Figure  \ref{fig:stability_LP} presents the same experiments with low-pass data. This reflects a phenomenon that typically occurs in optical applications in which the fine details of the data are blurred by the measurement process. Estimating a signal from its low-resolution measurements, when the phases are available, has been investigated thoroughly in the last years, see for instance~\cite{candes2014towards,bendory2017robust}. 
   Accordingly, we assume that we can acquire the data  $\mathbf{Z}[m,k]$ for all $m$ but only for  $k=-K_{\max},\dots,K_{\max}$ for some cut-off frequency $K_{\max}$. Particularly, in Figure  \ref{fig:stability_LP} we consider $N=53$ and $K_{\max}=18$ (i.e., $ 70\%$ of the spectral content) for the two proposed algorithms. In this case, if the SNR is not too bad, then NCPC works significantly better than ER in both cases.  As in Figure \ref{fig:stability}a, in the low SNR regime, minimizing the ER loss function achieves superior performance for Gaussian noise.

\section{\label{sec:theory}Theory}
This section presents the theoretical contribution of this work, focusing on the case of maximum overlap between adjacent windows $(L=1)$. As explained and demonstrated numerically, the non-convex approaches also tend to work well for $L>1$ and when the high-frequencies of the data are suppressed.

\begin{algorithm}
	\textbf{Input:} The measurements $\mathbf{Z}[m,k]$ as given in~\eqref{eq:meas0}
	and (optional) thresholding parameter $B>0$.
	
	\textbf{Output:} Estimation of \textbf{$\mathbf{x}$}. 
	\begin{enumerate}
		\item Initialization by Algorithm~\ref{alg:linearLS} (for $L=1$) or Algorithm~\ref{alg:linearLS_L} (for $L>1)$. 
		\item Apply  the update rule until convergence:
		
		\begin{enumerate}
			\item Gradient step:
			\[
			\tilde{\mathbf{x}}_{k}=\mathbf{x}_{k-1}-\mu\nabla f\left(\mathbf{x}_{k-1}\right),
			\]
			for step size $\mu$ and  $\nabla f$
			given  in~\eqref{eq:grad}. 
			\item Optional thresholding:
			\[
			\mathbf{x}_{k}[n]=\begin{cases}
			\tilde{\mathbf{x}}_{k}[n] & \mbox{if }\left|\tilde{\mathbf{x}}_{k}[n]\right|\leq B,\\
			B\cdot\operatorname{phase}\left(\tilde{\mathbf{x}}_{k}[n]\right) & \mbox{if }\left|\tilde{\mathbf{x}}_{k}[n]\right|>B.
			\end{cases}
			\]
			
		\end{enumerate}
	\end{enumerate}
	\protect\caption{\label{alg:5} Gradient descent algorithm to minimize the ER loss functions~\eqref{eq:LS_Z} or~\eqref{eq:LS}}
\end{algorithm}

In our first theoretical result, Theorem \ref{lem:LSinit}, we analyze
the initialization algorithm presented
in Algorithm \ref{alg:linearLS} and estimate the distance between the initialization vector and the ground truth. 
Next, we study the geometry of
the loss function~\eqref{eq:LS}, which controls the behavior of our ER minimization algorithm. To this end, suppose we minimize the ER loss function~\eqref{eq:LS} using gradient descent  followed by a thresholding step that can be used if the signal is bounded. This scheme is presented in Algorithm~\ref{alg:5}.
In Theorem~\ref{lem:converges} we establish the existence of a \emph{basin
of attraction} of size $\frac{1}{8\sqrt{N}W^{2}}$ around the global
minimum for signals with unit modulus entries. In the basin of attraction,
a gradient algorithm is guaranteed to converge to a global minimum at a
geometric rate. This result is true for any gradient scheme with a thresholding step as in Algorithm~\ref{alg:5}. 
  We stress that the theoretical contribution of this result is limited. As presented in Corollary \ref{thm:main}, the estimated basin of attraction is small so  that theoretically Algorithm~\ref{alg:5} converges in the same area in which the problem has a closed linear LS solution. To the best of our knowledge, this is the first result quantifying the size of the basin of attraction of a gradient algorithm in a deterministic phase retrieval setup. This is in contrast to the basin of attraction of  random phase retrieval setups which is quite well--understood. 

A crucial condition for the success of gradient algorithms is that its initialization is sufficiently close to the global minimum.
The following result quantifies the estimation error of the proposed
initialization presented in Algorithm \ref{alg:linearLS} for bounded signals and $L=1$. The error reduces to zero as $W$ approaches $\frac{N+1}{2}$.
The case of $L>1$ is discussed briefly in Section \ref{sec:Algorithm}. The result is stated for a normalized signal. The norm of the signal can be estimated easily from the main diagonal of $\mathbf{x}\mathbf{x}^*$ as explained in Section \ref{sec:uniqueness}.

\begin{thm}
\label{lem:LSinit}Suppose that $L=1$, $\|\mathbf{x}\|_2=1$, $\mathbf{g}$ is an admissible
window of length $W\geq 2$ and that $\|\mathbf{x}\|_{\infty}\leq\sqrt{\frac{B}{N}}$
for some $0<B\leq\frac{N}{2\left(N-2W+1\right)}$. Then under the
measurement model of (\ref{eq:meas0}), the initialization vector  given
in Algorithm \ref{alg:linearLS} satisfies 
\[
d^{2}\left(\mathbf{x}_{0},\mathbf{x}\right)\leq 2 \left(1-\sqrt{1-2B\frac{N-2W+1}{N}}\right).
\]
\end{thm}
\begin{IEEEproof}
See Section \ref{sub:Proof-of-Lemma_LSinit}. 
\end{IEEEproof}

The properties of the gradient algorithm minimizing the ER rely on the geometry of the loss function (\ref{eq:LS}) near the global minimum. 
The following result  quantifies the size of the basin of attraction of the loss function (\ref{eq:LS}),
namely, the area in which a gradient algorithm is guaranteed to converge
to a global minimum at a geometric rate. As demonstrated
in Figure \ref{fig:basin_of_attraction}, in practice the basin of
attraction is quite large 
for a broad family of signals. The proof relies on a geometric analysis of the loss function as presented in Lemmas \ref{lem:grad_bound} and \ref{lem:regualrity_con}.

\begin{thm}
\label{lem:converges}Let $L=1$ and suppose that $\mathbf{x\in\mathbb{R}}_{1/\sqrt{N}}^{N}$ and $\mathbf{g}$ is a rectangular window of length $W$.
Additionally, suppose that $d\left(\mathbf{x}_{0},\mathbf{x}\right)\leq\frac{1}{8\sqrt{N}W^{2}}$,
where $\mathbf{x}_{0}$ obeys $\left\Vert \mathbf{x}_{0}\right\Vert _{\infty}\leq\frac{1}{\sqrt{N}}$.
Then, under the measurement model (\ref{eq:meas0}), Algorithm \ref{alg:5}
with thresholding parameter $B=\frac{1}{\sqrt{N}}$ and step size $0<\mu\leq2/\beta$ achieves
the following geometric convergence: 
\[
d^{2}\left(\mathbf{x}_{k},\mathbf{x}\right)\leq\left(1-\frac{2\mu}{\alpha}\right)^{k}d^{2}\left(\mathbf{x}_{0},\mathbf{x}\right),
\]
where $\alpha  \geq  \frac{4N}{W}$ and $\beta  \geq  256N^2W^{3}$.
\end{thm}
\begin{IEEEproof}
See Section \ref{sub:Proof-of-Lemma-converges}. 
\end{IEEEproof}

Combining  Theorems \ref{lem:LSinit} and \ref{lem:converges} leads to the following corollary: 

\begin{corollary} \label{thm:main}Suppose that $L=1$, $\mathbf{x\in\mathbb{R}}_{1/\sqrt{N}}^{N}$, $N$ is a prime number and  $\mathbf{g}$ is a rectangular  window of length $W$ that satisfies: 
\[
2W-1+\frac{1}{128W^4}\geq N.
\]
 Then, under the measurement model of (\ref{eq:meas0}), 
Algorithm \ref{alg:5}, initialized by Algorithm \ref{alg:linearLS}, with  thresholding parameter $B=\frac{1}{\sqrt{N}}$ and step
size $0<\mu\leq2/\beta$ achieves the following geometric convergence:
\[
d^{2}\left(\mathbf{x}_{k},\mathbf{x}\right)\leq\left(1-\frac{2\mu}{\alpha}\right)^{k}d^{2}\left(\mathbf{x}_{0},\mathbf{x}\right),
\]
where  $\alpha  \geq  \frac{4N}{W}$ and $\beta  \geq  256N^2W^{3}$.
\end{corollary} 
\begin{IEEEproof}
See Section \ref{sec:Proof_MainTheorem}. 
\end{IEEEproof}

We mention that the result of Corollary \ref{thm:main} is good merely for long windows. However, in practice we observe that the algorithm works well also for short windows. As we discuss in Section \ref{sec:Conclusion}, bridging  this theoretical gap is an important direction for future research.


\section{\label{sec:Geometry}Proofs}


\subsection{\label{sub:Proof-of-Lemma_LSinit}Proof of Theorem \ref{lem:LSinit}}

 The initialization is based on extracting the principal eigenvector of the matrix $\mathbf{X}_0$ defined in Algorithm \ref{alg:linearLS}. 
 By assumption, $\mathbf{G}_{\ell}$
are invertible matrices for $\ell=-(W-1),\dots,W-1$ for some $W\geq 2$ and hence we
can compute (see (\ref{eq:Gl})) 
\[
\diag\left(\mathbf{X}_0,\ell\right)=\mathbf{G}_{\ell}^{-1}\mathbf{y}_{\ell}=\diag\left(\mathbf{X},\ell\right).
\]
For $\ell=W,\dots, N-W$ we have $\diag\left(\mathbf{X}_0,\ell\right)=0$.
Let us take a look at the matrix $\mathbf{E}:=\mathbf{X}-\mathbf{X}_{0}$.
Clearly, $\mathbf{E}$ is not zero at most on $N-2W+1$ diagonals.
In other words, in each row and column, there are at most $N-2W+1$
non-zero values. Let $\Omega_{i}$ be the set of non-zero values of
the $i${th} row of $\mathbf{E}$ with cardinality  $\left|\Omega_{i}\right|\leq N-2W+1$.
Using the fact that $\left\Vert \mathbf{x}\right\Vert _{\infty}=\sqrt{\frac{B}{{N}}}$
we can estimate 
\begin{eqnarray*}
\left\Vert \mathbf{E}\right\Vert _{\infty} & := & \max_{i}\sum_{j}\left|\mathbf{X}\left[i,j\right]-\mathbf{X}_{0}\left[i,j\right]\right|\\
 & = & \max_{i}\sum_{j\in\Omega_{i}}\left|\mathbf{X}\left[i,j\right]\right|\\
 & = & \max_{i}\sum_{j\in\Omega_{i}}\left|\mathbf{x}\left[i\right]\mathbf{x}\left[j\right]\right|\\
 & \leq & \frac{B\left(N-2W+1\right)}{N}.
\end{eqnarray*}
The same bound holds for $\left\Vert \mathbf{E}\right\Vert _{1}:=\max_{j}\sum_{i}\left|\mathbf{E}[i,j]\right|$
and therefore by H\"{o}lder's inequality we get 
\[
\left\Vert \mathbf{E}\right\Vert _{2}\leq\sqrt{\left\Vert \mathbf{E}\right\Vert _{\infty}\left\Vert \mathbf{E}\right\Vert _{1}}=\frac{B\left(N-2W+1\right)}{N}.
\]

In order to complete the proof, we still need to show
that if $\left\Vert \mathbf{X}-\mathbf{X}_{0}\right\Vert _{2}$ is
small, then $d\left(\mathbf{x},\mathbf{x}_{0}\right)$ is small as
well, where $\mathbf{{x}}_{0}$ is the principal eigenvector of $\mathbf{X}_{0}$ with appropriate normalization. To show that, we follow the outline of Section 7.8
in \cite{candes2015Wirtinger}. Observe that as $\mathbf{G}_{0}$ is invertible by assumption,  the norm of $\mathbf{x}$ is known by 
\begin{equation*}
\|\mathbf{x}\|_{2}^2=\sum_{n=0}^{N-1}\left(\diag\left(\mathbf{X},0\right)\right[n]=\sum_{n=0}^{N-1}\left({\mathbf{G}_{0}^{-1}\mathbf{y}_{0}}\right)[n].
\end{equation*}
Accordingly, we assume hereinafter without loss of generality that $\mathbf{x}$ and $\mathbf{x}_0$
have unit norm. 
Let $\lambda_{0}$ be the top eigenvalue
of $\mathbf{X}_{0}$, associated with $\mathbf{{x}}_{0}$. We observe
that 
\begin{eqnarray*}
\left|\lambda_{0}-\left|\mathbf{{x}}_{0}^{*}\mathbf{x}\right|^{2}\right| & = & \left|\mathbf{{x}}_{0}^{*}\mathbf{X}_{0}\mathbf{{x}}_{0}-\mathbf{{x}}_{0}^{*}\mathbf{x\mathbf{x}^{*}}\mathbf{{x}}_{0}\right|\\
 & \leq & \left\Vert \mathbf{X}_{0}-\mathbf{x\mathbf{x}^{*}}\right\Vert _{2}.
\end{eqnarray*}
Furthermore, as $\left\Vert \mathbf{x}\right\Vert _{2}=1$ we also
have 
\begin{eqnarray*}
\lambda_{0} & \geq & \mathbf{x}^{*}\mathbf{X}_{0}\mathbf{x}=\mathbf{x}^{*}\left(\mathbf{X}_{0}-\mathbf{x}\mathbf{x}^{*}\right)\mathbf{x}+1\\
 & \geq & 1-\left\Vert \mathbf{X}_{0}-\mathbf{x\mathbf{x}^{*}}\right\Vert _{2}.
\end{eqnarray*}
Combining the last two inequalities we get 
\begin{eqnarray*}
\left|\mathbf{x}_{0}^{*}\mathbf{x}\right|^{2} & \geq & 1-2\left\Vert \mathbf{X}_{0}-\mathbf{x\mathbf{x}^{*}}\right\Vert _{2}\\
 & \geq & 1-2B\frac{N-2W+1}{N}.
\end{eqnarray*}
It then follows that W
\begin{eqnarray*}
d^{2}\left(\mathbf{x}_{0},\mathbf{x}\right) & \leq & 2\left(1-\sqrt{1-2B\frac{N-2W+1}{N}}\right),
\end{eqnarray*}
where the term in the square root is positive by assumption.


\subsection{\label{sub:Proof-of-Lemma-converges}Proof of Theorem \ref{lem:converges}}

For fixed $\mathbf{x}$, let $\mathcal{E}$ be the set of vectors
in $\mathbb{R}^{N}$ satisfying $\mathbf{\left\Vert z\right\Vert _{\infty}}\leq\frac{1}{\sqrt{N}}$
and $d\left(\mathbf{x},\mathbf{z}\right)\leq\frac{1}{8\sqrt{N}W^{2}}$.
We first need the following definition: 
\begin{defn}
\label{def:regularity-1}We say that a function $f$ satisfies the
\emph{regularity condition} in $\mathcal{E}$ if for all vectors $\mathbf{z}\in\mathcal{E}$
we have 
\[
\left\langle \nabla f(\mathbf{z}),\mathbf{z}-\mathbf{x}e^{j\phi(\mathbf{z})}\right\rangle \geq\frac{1}{\alpha}d^{2}\left(\mathbf{z},\mathbf{x}\right)+\frac{1}{\beta}\left\Vert \nabla f(\mathbf{z})\right\Vert _{2}^{2},
\]
for some  positive constants $\alpha,\beta$. 
\end{defn}
The following lemma states that if the regularity condition is met, then
the gradient step converges to a global minimum at a geometric rate. 
\begin{lem} \label{lem:regularity}
Assume that $f$ satisfies the regularity condition for all $\mathbf{z}\in\mathcal{E}$.
Consider the following update rule 
\[
\mathbf{z}_{k}=\mathbf{z}_{k-1}-\mu\nabla f\left(\mathbf{z}_{k-1}\right),
\]
for $0<\mu\leq2/\beta$. Then, 
\[
d^{2}\left(\mathbf{z}_{k},\mathbf{x}\right)\leq\left(1-\frac{2\mu}{\alpha}\right)^{}d^{2}\left(\mathbf{z}_{k-1},\mathbf{x}\right).
\]
\end{lem}
\begin{IEEEproof}
See Section 7.4 in \cite{candes2015Wirtinger}. 
\end{IEEEproof}

In order to show that the regularity condition of Definition \ref{def:regularity-1} is met, we present two lemmas 
for signals with unit modulus entries. 
The first result shows that the gradient of the loss function (\ref{eq:LS}),
given explicitly in (\ref{eq:grad}), is bounded near its global minimum. This
implies that the loss function is smooth. We consider here only the
case of a rectangular window $\mathbf{g}$ of length $W$. The extension
to non-vanishing windows of length $W$ is straightforward (see remark in Appendix \ref{sub:Proof-of-Lemma_grad_bound}): 
\begin{lem}
\label{lem:grad_bound}Suppose that $\mathbf{x}\in\mathbb{R}_{1/\sqrt{N}}^{N}$,
$\left\Vert \mathbf{z}\right\Vert _{\infty}\leq\frac{1}{\sqrt{N}}$
and $d\left(\mathbf{x},\mathbf{z}\right)\leq\frac{1}{\sqrt{N}}$.
Let $\mathbf{g}$ be a rectangular window of length $W$. Then, $\nabla f(\mathbf{z})$ as 
given in (\ref{eq:grad}) satisfies 
\[
\left\Vert \nabla f(\mathbf{z})\right\Vert _{2}\leq\frac{8}{L}W^{2}\sqrt{N}d(\mathbf{x},\mathbf{z}).
\]
\end{lem}
\begin{IEEEproof}
See Appendix \ref{sub:Proof-of-Lemma_grad_bound}. 
\end{IEEEproof}
The second lemma shows that the inner product between the gradient
and the vector $\mathbf{z}-\mathbf{x}e^{j\phi(\mathbf{z})}$ is positive
if $d\left(\mathbf{x},\mathbf{z}\right)\leq\frac{1}{8\sqrt{N}W^{2}}$.
This result implies that $-\nabla f(\mathbf{z})$ points approximately 
towards $\mathbf{x}$. As in Lemma \ref{lem:grad_bound}, we consider for simplicity 
 rectangular windows of length $W$. Yet, the analysis can 
be extended to non-vanishing windows of length $W$. In this case,
the bounds are dependent on the dynamic range of
$\mathbf{g}$ (for details, see remark in Appendix  \ref{rem:inner_prod}). 
\begin{lem}
\label{lem:regualrity_con} Suppose that $L=1$ and $\mathbf{g}$
is a rectangular window of length $W$. For any $\mathbf{x}\in\mathbb{R}_{1/\sqrt{N}}^{N}$
and $\mathbf{\left\Vert z\right\Vert _{\infty}}\leq\frac{1}{\sqrt{N}}$, if $d\left(\mathbf{x},\mathbf{z}\right)\leq\frac{1}{8\sqrt{N}W^{2}}$,
then 
\[
\left\langle \nabla f(\mathbf{z}),\mathbf{z}-\mathbf{x}e^{j\phi(\mathbf{z)}}\right\rangle \geq\frac{Wd^{2}\left(\mathbf{x},\mathbf{z}\right)}{2N},
\]
where $\nabla f(\mathbf{z})$ is given in (\ref{eq:grad}). \end{lem}
\begin{IEEEproof}
See Appendix \ref{sub:Proof-of-Lemma_regularity_con}. 
\end{IEEEproof}

We notice that the thresholding stage of Algorithm \ref{alg:5} cannot increase the error as the signal is assumed to be bounded. 
The proof of Theorem \ref{lem:converges} is then completed by  directly leveraging lemmas \ref{lem:grad_bound} and \ref{lem:regualrity_con}
and seeing that Definition \ref{def:regularity-1} holds in our case
with  constants $\alpha  \geq \frac{4N}{W}$ and  $\beta \geq  256N^2W^{3}$.


\subsection{\label{sec:Proof_MainTheorem}Proof of Corollary \ref{thm:main}}
As $N$ is a prime number, $\mathbf{g}$ is an admissible window of length $W$ (see Lemma~\ref{claim:rect_win}). 
According to Theorem \ref{lem:converges}, we merely need to show that
the initialization point is within the basin of attraction, namely,
$d\left(\mathbf{x},\mathbf{x}_{0}\right)\leq\frac{1}{8\sqrt{N}W^{2}}$.
From Lemma~\ref{lem:LSinit}, we know that the initialization
obeys 
\begin{eqnarray*}
d^{2}\left(\mathbf{x}_{0},\mathbf{x}\right) & \leq & 2\left(1-\sqrt{1-2\frac{N-2W+1}{N}}\right).
\end{eqnarray*}
Using the fact that $a\leq\sqrt{a}$ for all $0\leq a\leq1$ and some
standard algebraic calculations, we conclude that the initialization
of Algorithm \ref{alg:linearLS} is within the basin of attraction
as long as 
\[
2W-1+\frac{1}{128W^{4}}\geq N,
\]
which completes the proof.


\section{\label{sec:Conclusion}Discussion}
 
This  paper explores  practical, efficient, non-convex 
phase retrieval algorithms with some deterministic theoretical guarantees.
Particularly, we propose two local optimization methods based on minimizing the ER loss function and optimizing on the manifold of phases. The latter is a new phase retrieval
algorithm that takes into account the special geometry of the phase retrieval problem. 

Since the optimization problems are non-convex, we also propose an initialization method. The method is based on the insight that, for sufficiently long windows, the signal can be recovered as the solution of a linear LS problem. While this may not be true for shorter windows, we use the LS solution to construct a special matrix and initialize the local optimization algorithms with the principal eigenvector of this matrix.  Similar initialization approaches were suggested recently
for phase retrieval problems. However, they are mainly focused on
random setups and  based
on {probabilistic} considerations.
For $L=1$, we estimate the distance between the initialization point and the ground truth.  
 The case of $L>1$ raises some interesting
questions. As a heuristic, we suggested to smoothly
interpolate the missing entries. This practice works quite well since
the window acts as an averaging operator. Clearly, the interpolation
method depends on the window shape. 
A main challenge for future research is analyzing  the setting of $L>1$. 

For signals with unit
modulus entries, we prove in Theorem \ref{lem:converges} that the ER loss function  has a basin of attraction.  We show numerically that the actual basin of attraction is larger than the theoretical bound and exists for a broader family of signals.
The gap between the actual size of the basin of attraction and the theoretical result is the bottleneck that prevents a full theoretical understanding of the proposed algorithms. Specifically, improving  Lemma \ref{lem:regualrity_con} will lead directly to tighter estimation of the size of the basin of attraction. Ideally, this would lead to the conclusion that the proposed initial guess lies in the basin.


\section*{acknowledgment}

We would like to thank Mahdi Soltanolkotabi, {Ir\'ene Waldspurger, Pavel Sidorenko and Laura Waller} for their remarks on an initial draft of this paper.


\bibliographystyle{ieeetrans}
\bibliography{refs}

\begin{thebibliography}{10}
\providecommand{\url}[1]{#1}
\csname url@samestyle\endcsname
\providecommand{\newblock}{\relax}
\providecommand{\bibinfo}[2]{#2}
\providecommand{\BIBentrySTDinterwordspacing}{\spaceskip=0pt\relax}
\providecommand{\BIBentryALTinterwordstretchfactor}{4}
\providecommand{\BIBentryALTinterwordspacing}{\spaceskip=\fontdimen2\font plus
\BIBentryALTinterwordstretchfactor\fontdimen3\font minus
  \fontdimen4\font\relax}
\providecommand{\BIBforeignlanguage}[2]{{%
\expandafter\ifx\csname l@#1\endcsname\relax
\typeout{** WARNING: IEEEtranS.bst: No hyphenation pattern has been}%
\typeout{** loaded for the language `#1'. Using the pattern for}%
\typeout{** the default language instead.}%
\else
\language=\csname l@#1\endcsname
\fi
#2}}
\providecommand{\BIBdecl}{\relax}
\BIBdecl

\bibitem{genrtr}
P.-A. Absil, C.~G. Baker, and K.~A. Gallivan, ``Trust-region methods on
  {Riemannian} manifolds,'' \emph{Foundations of Computational Mathematics},
  vol.~7, no.~3, pp. 303--330, 2007.

\bibitem{AMS08}
P.-A. Absil, R.~Mahony, and R.~Sepulchre, \emph{Optimization Algorithms on
  Matrix Manifolds}.\hskip 1em plus 0.5em minus 0.4em\relax Princeton, NJ:
  Princeton University Press, 2008.

\bibitem{bandeira2014tightness}
A.~S. Bandeira, N.~Boumal, and A.~Singer, ``Tightness of the maximum likelihood
  semidefinite relaxation for angular synchronization,'' \emph{Mathematical
  Programming}, pp. 1--23, 2016.

\bibitem{bandeira2015non}
A.~S. Bandeira, Y.~Chen, and A.~Singer, ``Non-unique games over compact groups
  and orientation estimation in cryo-em,'' \emph{arXiv preprint
  arXiv:1505.03840}, 2015.

\bibitem{baykal2004blind}
B.~Baykal, ``Blind channel estimation via combining autocorrelation and blind
  phase estimation,'' \emph{Circuits and Systems I: Regular Papers, IEEE
  Transactions on}, vol.~51, no.~6, pp. 1125--1131, 2004.

\bibitem{bendory2017uniqueness}
T.~Bendory, P.~Sidorenko, and Y.~C. Eldar, ``On the uniqueness of {FROG}
  methods,'' \emph{IEEE Signal Processing Letters}, vol.~24, no.~5, pp.
  722--726, 2017.

\bibitem{bendory2017robust}
T.~Bendory, ``Robust recovery of positive stream of pulses,'' \emph{IEEE
  Transactions on Signal Processing}, vol.~65, no.~8, pp. 2114--2122, 2017.

\bibitem{bendory2017fourier}
T.~Bendory, R.~Beinert, and Y.~C. Eldar, ``Fourier phase retrieval: Uniqueness
  and algorithms,'' \emph{arXiv preprint arXiv:1705.09590}, 2017.

\bibitem{bendory2017bispectrum}
T.~Bendory, N.~Boumal, C.~Ma, Z.~Zhao, and A.~Singer, ``Bispectrum inversion
  with application to multireference alignment,'' \emph{arXiv preprint
  arXiv:1705.00641}, 2017.

\bibitem{bendory2017signal}
T.~Bendory, D.~Edidin, and Y.~C. Eldar, ``On signal reconstruction from frog
  measurements,'' \emph{arXiv preprint arXiv:1706.08494}, 2017.

\bibitem{bojarovska2015phase}
I.~Bojarovska and A.~Flinth, ``Phase retrieval from gabor measurements,''
  \emph{Journal of Fourier Analysis and Applications}, vol.~22, no.~3, pp.
  542--567, 2016.

\bibitem{boumal2016globalrates}
N.~Boumal, P.-A. Absil, and C.~Cartis, ``Global rates of convergence for
  nonconvex optimization on manifolds,'' \emph{arXiv preprint
  arXiv:1605.08101}, 2016.

\bibitem{manopt}
\BIBentryALTinterwordspacing
N.~Boumal, B.~Mishra, P.-A. Absil, and R.~Sepulchre, ``{M}anopt, a {M}atlab
  toolbox for optimization on manifolds,'' \emph{Journal of Machine Learning
  Research}, vol.~15, pp. 1455--1459, 2014. [Online]. Available:
  \url{http://www.manopt.org}
\BIBentrySTDinterwordspacing

\bibitem{boumal2016nonconvex}
N.~Boumal, ``Nonconvex phase synchronization,'' \emph{SIAM Journal on
  Optimization}, vol.~26, no.~4, pp. 2355--2377, 2016.

\bibitem{candes2015phase}
E.~Candes, Y.~C. Eldar, T.~Strohmer, and V.~Voroninski, ``Phase retrieval via
  matrix completion,'' \emph{SIAM Review}, vol.~57, no.~2, pp. 225--251, 2015.

\bibitem{candes2014solving}
E.~Cand{\`e}s and X.~Li, ``Solving quadratic equations via {P}haselift when
  there are about as many equations as unknowns,'' \emph{Foundations of
  Computational Mathematics}, vol.~14, no.~5, pp. 1017--1026, 2014.

\bibitem{candes2014phase}
E.~Candes, X.~Li, and M.~Soltanolkotabi, ``Phase retrieval from coded
  diffraction patterns,'' \emph{Applied and Computational Harmonic Analysis},
  vol.~39, no.~2, pp. 277--299, 2015.

\bibitem{candes2015Wirtinger}
------, ``Phase retrieval via {W}irtinger flow: Theory and algorithms,''
  \emph{Information Theory, IEEE Transactions on}, vol.~61, no.~4, pp.
  1985--2007, 2015.

\bibitem{candes2013phaselift}
E.~Candes, T.~Strohmer, and V.~Voroninski, ``Phaselift: Exact and stable signal
  recovery from magnitude measurements via convex programming,''
  \emph{Communications on Pure and Applied Mathematics}, vol.~66, no.~8, pp.
  1241--1274, 2013.

\bibitem{candes2014towards}
E.~J. Cand{\`e}s and C.~Fernandez-Granda, ``Towards a mathematical theory of
  super-resolution,'' \emph{Communications on Pure and Applied Mathematics},
  vol.~67, no.~6, pp. 906--956, 2014.

\bibitem{chen2016projected}
Y.~Chen and E.~Candes, ``The projected power method: An efficient algorithm for
  joint alignment from pairwise differences,'' \emph{arXiv preprint
  arXiv:1609.05820}, 2016.

\bibitem{chen2015solving}
Y.~Chen and E.~J. Candes, ``Solving random quadratic systems of equations is
  nearly as easy as solving linear systems,'' \emph{Communications on Pure and
  Applied Mathematics}, vol.~70, no.~5, pp. 822--883, 2017.

\bibitem{delong1994frequency}
K.~DeLong, R.~Trebino, J.~Hunter, and W.~White, ``Frequency-resolved optical
  gating with the use of second-harmonic generation,'' \emph{JOSA B}, vol.~11,
  no.~11, pp. 2206--2215, 1994.

\bibitem{dumitrescu2007positive}
B.~Dumitrescu, \emph{Positive trigonometric polynomials and signal processing
  applications}.\hskip 1em plus 0.5em minus 0.4em\relax Springer Science \&
  Business Media, 2007.

\bibitem{eldar2014phase}
Y.~C. Eldar and S.~Mendelson, ``Phase retrieval: Stability and recovery
  guarantees,'' \emph{Applied and Computational Harmonic Analysis}, vol.~36,
  no.~3, pp. 473--494, 2014.

\bibitem{eldar2015sparse}
Y.~C. Eldar, P.~Sidorenko, D.~Mixon, S.~Barel, and O.~Cohen, ``Sparse phase
  retrieval from short-time fourier measurements,'' \emph{Signal Processing
  Letters, IEEE}, vol.~22, no.~5, pp. 638--642, 2015.

\bibitem{fienup1987phase}
C.~Fienup and J.~Dainty, ``Phase retrieval and image reconstruction for
  astronomy,'' \emph{Image Recovery: Theory and Application}, pp. 231--275,
  1987.

\bibitem{fienup1982phase}
J.~Fienup, ``Phase retrieval algorithms: a comparison,'' \emph{Applied optics},
  vol.~21, no.~15, pp. 2758--2769, 1982.

\bibitem{gerchberg1972practical}
R.~Gerchberg and W.~Saxton, ``A practical algorithm for the determination of
  phase from image and diffraction plane pictures,'' \emph{Optik}, vol.~35, p.
  237, 1972.

\bibitem{goemans1995improved}
M.~Goemans and D.~Williamson, ``Improved approximation algorithms for maximum
  cut and satisfiability problems using semidefinite programming,''
  \emph{Journal of the ACM (JACM)}, vol.~42, no.~6, pp. 1115--1145, 1995.

\bibitem{griffin1984signal}
D.~Griffin and J.~Lim, ``Signal estimation from modified short-time fourier
  transform,'' \emph{Acoustics, Speech and Signal Processing, IEEE Transactions
  on}, vol.~32, no.~2, pp. 236--243, 1984.

\bibitem{gross2014improved}
D.~Gross, F.~Krahmer, and R.~Kueng, ``Improved recovery guarantees for phase
  retrieval from coded diffraction patterns,'' \emph{Applied and Computational
  Harmonic Analysis}, 2015.

\bibitem{harrison1993phase}
R.~Harrison, ``Phase problem in crystallography,'' \emph{JOSA A}, vol.~10,
  no.~5, pp. 1046--1055, 1993.

\bibitem{Huang2015}
K.~Huang, Y.~C. Eldar, and N.~D. Sidiropoulos, ``Phase retrieval from 1d
  fourier measurements: Convexity, uniqueness, and algorithms,'' \emph{IEEE
  Transactions on Signal Processing}, vol.~64, no.~23, pp. 6105--6117, 2016.

\bibitem{iwen2016phase}
M.~A. Iwen, B.~Preskitt, R.~Saab, and A.~Viswanathan, ``Phase retrieval from
  local measurements: Improved robustness via eigenvector-based angular
  synchronization,'' \emph{arXiv preprint arXiv:1612.01182}, 2016.

\bibitem{iwen2016fast}
M.~A. Iwen, A.~Viswanathan, and Y.~Wang, ``Fast phase retrieval from local
  correlation measurements,'' \emph{SIAM Journal on Imaging Sciences}, vol.~9,
  no.~4, pp. 1655--1688, 2016.

\bibitem{jaganathan2015phase}
K.~Jaganathan, Y.~C. Eldar, and B.~Hassibi, ``Phase retrieval: An overview of
  recent developments,'' \emph{arXiv preprint arXiv:1510.07713}, 2015.

\bibitem{jaganathan2015stft}
------, ``{STFT} phase retrieval: Uniqueness guarantees and recovery
  algorithms,'' \emph{IEEE Journal of Selected Topics in Signal Processing},
  vol.~10, no.~4, pp. 770--781, 2016.

\bibitem{jaganathan2013sparse}
K.~Jaganathan, S.~Oymak, and B.~Hassibi, ``Sparse phase retrieval: Convex
  algorithms and limitations,'' in \emph{Information Theory Proceedings (ISIT),
  2013 IEEE International Symposium on}.\hskip 1em plus 0.5em minus 0.4em\relax
  IEEE, 2013, pp. 1022--1026.

\bibitem{journee2010generalized}
M.~Journ{\'e}e, Y.~Nesterov, P.~Richt{\'a}rik, and R.~Sepulchre, ``Generalized
  power method for sparse principal component analysis,'' \emph{The Journal of
  Machine Learning Research}, vol.~11, pp. 517--553, 2010.

\bibitem{juang1993fundamentals}
B.~Juang and L.~Rabiner, ``Fundamentals of speech recognition,'' \emph{Signal
  Processing Series. Prentice Hall, Englewood Cliffs, NJ}, 1993.

\bibitem{kreutz2009complex}
K.~Kreutz-Delgado, ``The complex gradient operator and the cr-calculus,''
  \emph{arXiv preprint arXiv:0906.4835}, 2009.

\bibitem{lee2016gradient}
J.~Lee, M.~Simchowitz, M.~Jordan, and B.~Recht, ``Gradient descent converges to
  minimizers,'' \emph{arXiv preprint arXiv:1602.04915}, 2016.

\bibitem{maiden2011superresolution}
A.~Maiden, M.~Humphry, F.~Zhang, and J.~Rodenburg, ``Superresolution imaging
  via ptychography,'' \emph{JOSA A}, vol.~28, no.~4, pp. 604--612, 2011.

\bibitem{marchesini2015alternating}
S.~Marchesini, Y.~Tu, and H.~Wu, ``Alternating projection, ptychographic
  imaging and phase synchronization,'' \emph{Applied and Computational Harmonic
  Analysis}, 2015.

\bibitem{millane1990phase}
R.~Millane, ``Phase retrieval in crystallography and optics,'' \emph{JOSA A},
  vol.~7, no.~3, pp. 394--411, 1990.

\bibitem{nawab1983signal}
S.~Nawab, T.~Quatieri, and J.~Lim, ``Signal reconstruction from short-time
  fourier transform magnitude,'' \emph{Acoustics, Speech and Signal Processing,
  IEEE Transactions on}, vol.~31, no.~4, pp. 986--998, 1983.

\bibitem{netrapalli2015phase}
P.~Netrapalli, P.~Jain, and S.~Sanghavi, ``Phase retrieval using alternating
  minimization,'' \emph{Signal Processing, IEEE Transactions on}, vol.~63,
  no.~18, pp. 4814--4826, 2015.

\bibitem{oppenheim2010discrete}
A.~Oppenheim and R.~Schafer, \emph{Discrete-time signal processing}.\hskip 1em
  plus 0.5em minus 0.4em\relax Pearson Higher Education, 2010.

\bibitem{pauwels2017fienup}
E.~Pauwels, A.~Beck, Y.~C. Eldar, and S.~Sabach, ``On {F}ienup methods for
  regularized phase retrieval,'' \emph{arXiv preprint arXiv:1702.08339}, 2017.

\bibitem{perry2016message}
A.~Perry, A.~S. Wein, A.~S. Bandeira, and A.~Moitra, ``Message-passing
  algorithms for synchronization problems over compact groups,'' \emph{arXiv
  preprint arXiv:1610.04583}, 2016.

\bibitem{pfander2016robust}
G.~E. Pfander and P.~Salanevich, ``Robust phase retrieval algorithm for
  time-frequency structured measurements,'' \emph{arXiv preprint
  arXiv:1611.02540}, 2016.

\bibitem{ranieri2013phase}
J.~Ranieri, A.~Chebira, Y.~M. Lu, and M.~Vetterli, ``Phase retrieval for sparse
  signals: Uniqueness conditions,'' \emph{arXiv preprint arXiv:1308.3058},
  2013.

\bibitem{rodenburg2008ptychography}
J.~Rodenburg, ``Ptychography and related diffractive imaging methods,''
  \emph{Advances in Imaging and Electron Physics}, vol. 150, no.~07, pp.
  87--184, 2008.

\bibitem{rodenburg1992theory}
J.~Rodenburg and R.~Bates, ``The theory of super-resolution electron microscopy
  via wigner-distribution deconvolution,'' \emph{Philosophical Transactions of
  the Royal Society of London A: Mathematical, Physical and Engineering
  Sciences}, vol. 339, no. 1655, pp. 521--553, 1992.

\bibitem{rusu2007extending}
C.~Rusu and J.~Astola, ``Extending a sequence into a minimum-phase sequence,''
  in \emph{In: Bregovic, R. \& Gotchev, A.(eds.). Proceedings of the 2007
  International TICSP Workshop on Spectral Methods and Multirate Signal
  Processing, SMMSP 2007, Moscow, Russia, 1-2 September 2007}, 2007.

\bibitem{sanghavi2015local}
S.~Sanghavi, R.~Ward, and C.~D. White, ``The local convexity of solving systems
  of quadratic equations,'' \emph{Results in Mathematics}, pp. 1--40, 2016.

\bibitem{shechtman2014gespar}
Y.~Shechtman, A.~Beck, and Y.~C. Eldar, ``{GESPAR}: Efficient phase retrieval
  of sparse signals,'' \emph{Signal Processing, IEEE Transactions on}, vol.~62,
  no.~4, pp. 928--938, 2014.

\bibitem{shechtman2014phase}
Y.~Shechtman, Y.~C. Eldar, O.~Cohen, H.~Chapman, J.~Miao, and M.~Segev, ``Phase
  retrieval with application to optical imaging: a contemporary overview,''
  \emph{Signal Processing Magazine, IEEE}, vol.~32, no.~3, pp. 87--109, 2015.

\bibitem{shechtman2011sparsity}
Y.~Shechtman, Y.~C. Eldar, A.~Szameit, and M.~Segev, ``Sparsity based
  sub-wavelength imaging with partially incoherent light via quadratic
  compressed sensing,'' \emph{Optics express}, vol.~19, no.~16, pp.
  14\,807--14\,822, 2011.

\bibitem{singer2011angular}
A.~Singer, ``Angular synchronization by eigenvectors and semidefinite
  programming,'' \emph{Applied and computational harmonic analysis}, vol.~30,
  no.~1, pp. 20--36, 2011.

\bibitem{sun2016geometric}
J.~Sun, Q.~Qu, and J.~Wright, ``A geometric analysis of phase retrieval,''
  \emph{arXiv preprint arXiv:1602.06664}, 2016.

\bibitem{trebino2002frequency}
R.~Trebino, \emph{Frequency-resolved optical gating: the measurement of
  ultrashort laser pulses}.\hskip 1em plus 0.5em minus 0.4em\relax Springer
  Science \& Business Media, 2012.

\bibitem{waldspurger2015phase}
I.~Waldspurger, A.~dAspremont, and S.~Mallat, ``Phase recovery, maxcut and
  complex semidefinite programming,'' \emph{Mathematical Programming}, vol.
  149, no. 1-2, pp. 47--81, 2015.

\bibitem{waldspurger2016phase}
I.~Waldspurger, ``Phase retrieval with random gaussian sensing vectors by
  alternating projections,'' \emph{arXiv preprint arXiv:1609.03088}, 2016.

\bibitem{walther1963question}
A.~Walther, ``The question of phase retrieval in optics,'' \emph{Journal of
  Modern Optics}, vol.~10, no.~1, pp. 41--49, 1963.

\bibitem{wang2016solving}
G.~Wang, G.~Giannakis, and Y.~C. Eldar, ``Solving systems of random quadratic
  equations via truncated amplitude flow,'' \emph{arXiv preprint
  arXiv:1605.08285}, 2016.

\bibitem{wang2014phase}
Y.~Wang and Z.~Xu, ``Phase retrieval for sparse signals,'' \emph{Applied and
  Computational Harmonic Analysis}, vol.~37, no.~3, pp. 531--544, 2014.

\bibitem{yang2011iterative}
C.~Yang, J.~Qian, A.~Schirotzek, F.~Maia, and S.~Marchesini, ``Iterative
  algorithms for ptychographic phase retrieval,'' \emph{arXiv preprint
  arXiv:1105.5628}, 2011.

\bibitem{yeh2015experimental}
L.-H. Yeh, J.~Dong, J.~Zhong, L.~Tian, M.~Chen, G.~Tang, M.~Soltanolkotabi, and
  L.~Waller, ``Experimental robustness of fourier ptychography phase retrieval
  algorithms,'' \emph{Optics express}, vol.~23, no.~26, pp. 33\,214--33\,240,
  2015.

\bibitem{zhang2016reshaped}
H.~Zhang and Y.~Liang, ``Reshaped wirtinger flow for solving quadratic system
  of equations,'' in \emph{Advances in Neural Information Processing Systems},
  2016, pp. 2622--2630.

\end{thebibliography}


\appendix


\subsection{\label{sec:algebriac_alg}Proof of Proposition \ref{thm:algebriac}}
 By assumption, the DFT of $\mathbf{g}\odot\left(\mathbf{P}_{-\ell}\mathbf{g}\right)$
is non-vanishing for $\ell=0,1,$ and the matrices $\mathbf{G}_{\ell},\thinspace\ell=0,1$
as given in (\ref{eq:Gl}) are invertible. Then,  we can compute
\[
\mathbf{x}_{\ell}=\mathbf{G}_{\ell}^{-1}\mathbf{y}_{\ell},\quad\ell=0,1,
\]
where $\mathbf{X}=\mathbf{x}\mathbf{x}^*$, $\mathbf{x}_{\ell}=\diag\left(\mathbf{X},\ell\right)$
and $\mathbf{y}_{\ell}:=\left\{ \mathbf{Y}\left[m,\ell\right]\right\} _{m=0}^{N-1}$.
Because of the fundamental ambiguity of phase retrieval, the first
entry can be set arbitrarily to  $\sqrt{\mathbf{x}_{0}\left[0\right]}=\left|\mathbf{x}\left[0\right]\right|$.
Then, as we assume non-vanishing signals, the rest of the entries
 are determined recursively for $n=1\dots,N-1$ by 
\[
\frac{\mathbf{x}_{1}\left[n-1\right]}{{\mathbf{x}}\left[n-1\right]}=\frac{\mathbf{x}\left[n-1\right]\mathbf{x}\left[n\right]}{{\mathbf{x}}\left[n-1\right]}=\mathbf{x}\left[n\right].
\]
This completes the proof.


\subsection{ \label{proof_of_prop_unit_signal}  Proof of Proposition \ref{cor:init_unit_signal}}
By assumption, $\mathbf{G}_\ell$ is an invertible matrix for $\vert \ell\vert \leq W-1$ for some $W\geq 2$ (see (\ref{eq:Gl})). Hence, we can compute $\diag\left(\mathbf{X},\ell\right)=\mathbf{G}_\ell^{-1}\mathbf{y}_\ell$ for $\ell=0,M$ for any $1\leq M\leq W-1$. The proof is a direct corollary of the following lemma:
\begin{lem*}
\label{lem:spectral_init} Let $L=1$. Suppose that $\mathbf{x}\in\mathbb{C}_{1/\sqrt{N}}^{N}$
and let $\mathbf{X}=\mathbf{x}\mathbf{x}^{*}$. Fix $M\in\{1,\dots, N-1\}$ and 
let $\mathbf{X}_{0}$ be a matrix obeying 
\begin{equation*}
\diag\left(\mathbf{X}_{0},\ell\right)=\begin{cases}
\diag\left(\mathbf{X},\ell\right), & \quad\ell=0,M,\\
0, & \quad\mbox{otherwise}.
\end{cases} 
\end{equation*}
Then, $\mathbf{x}$ is a principal eigenvectors of $\mathbf{X}_{0}$
(up to global phase). 
\end{lem*}

\begin{IEEEproof}
Based on the special structure of $\mathbf{X_0}$, the following calculation shows that $\mathbf{x}$ is an eigenvector
of $\mathbf{X}_{0}$ with $\frac{2}{N}$ as the associated eigenvalue:
\begin{eqnarray*}
\left(\mathbf{X}_{0}\mathbf{x}\right)[i] & = & \sum_{j=1}^{N}\mathbf{X}_{0}[i,j]\mathbf{x}[j]\\
 & = & \mathbf{X}_{0}[i,i]\mathbf{x}[i]+\mathbf{X}_{0}[i,i+M]\mathbf{x}[i+M]\\
 & = & \mathbf{x}[i]\left|\mathbf{x}[i]\right|^{2}+\mathbf{x}[i]\left|\mathbf{x}[i+M]\right|^{2}\\
 & = & \frac{2}{N}\mathbf{x}[i].
\end{eqnarray*}

We still need to show that $\mathbf{x}$ is a principal eigenvector of $\mathbf{X}_0$. Since each column and row of $\mathbf{X}_{0}$ is composed of two
non-zero values, it is evident that 
\[
\left\Vert \mathbf{X}_{0}\right\Vert _{\infty}:=\max_{i}\sum_{j}\left|\mathbf{X}_{0}\left[i,j\right]\right|=\frac{2}{N}.
\]
In the same manner \[ 
\left\Vert \mathbf{X}_{0}\right\Vert _{1}:= \max_{j}\sum_{i}\left|\mathbf{X}_{0}\left[i,j\right]\right|=\frac{2}{N}. \]
Hence by H\"{o}lder inequality we get

\[
\left\Vert \mathbf{X}_{0}\right\Vert _{2}\leq\sqrt{\left\Vert \mathbf{X}_{0}\right\Vert _{1}\left\Vert \mathbf{X}_{0}\right\Vert _{\infty}}=\frac{2}{N}.
\]
completing the proof.
\end{IEEEproof}


\subsection{\label{sec:linearLS}Proof of Proposition \ref{thm:linearLS}}

As the matrices $\mathbf{G}_\ell$ are invertible by assumption for all $\ell=-(W-1),\dots,(W-1)$, we can compute  
\[\diag\left(\mathbf{X}_{0},\ell\right)=\mathbf{G}_{\ell}^{-1}\mathbf{y}_{\ell} =\diag\left(\mathbf{X},\ell\right). \] The assumption $W\geq  \left\lceil \frac{N+1}{2}\right\rceil$ implies that $\mathbf{X}_0=\mathbf{X}$. Specifically, observe that it is sufficient to consider only $W= \left\lceil \frac{N+1}{2}\right\rceil$ since
 for any $\vert \ell_1\vert>\left \lceil\frac{N+1}{2}\right\rceil$, the window $\mathbf{g}\odot\left(\mathbf{P}_{-\ell_1}\right)$ is equal to another window $\mathbf{g}\odot\left(\mathbf{P}_{-\ell_2}\right)$ for some $\vert \ell_2\vert\leq \left\lceil\frac{N+1}{2}\right\rceil$. 
 
Let $\mathbf{\tilde{x}}:=\mathbf{x}/\|\mathbf{x}\|_2$. 
Then,  $\mathbf{\tilde{x}}$ is the principal eigenvector of $\mathbf{X}$ and the normalization stage of  Algorithm \ref{alg:linearLS} gives 
\[ 
\sqrt{\sum_{n=0}^{N-1}\left(\mathbf{G}_0^{-1}\mathbf{y}_0\right)[n]}=\|\mathbf{x}\|_2.
\]


\subsection{\label{sec:proof_of_equivalence}Proof of the equality between the loss functions~(\ref{eq:LS}) and~(\ref{eq:LS_Z})}

Recall that 
\begin{equation*}
\begin{split}
f(\mathbf{u})&=\frac{1}{2}\sum_{m=0}^{\left\lceil \frac{N}{L} \right\rceil-1}\sum_{k=0}^{N-1}\left(\mathbf{u}^{*}\mathbf{\tilde{H}}_{m,k}\mathbf{u}-\mathbf{Z}[m,k]\right)^{2} \\
&=\frac{1}{2}\sum_{m=0}^{\left\lceil \frac{N}{L} \right\rceil-1} \| \mathbf{\tilde{H}}_m- \mathbf{Z}_m\|_2^2,
\end{split}
\end{equation*}
where $\mathbf{Z}_m:=\{\mathbf{Z}[m,k]\}_{k=0}^{N-1}\in\mathbb{R}^N$ and $\tilde{\mathbf{H}}_m:=\{\mathbf{u}^{*}\mathbf{\tilde{H}}_{m,k}\mathbf{u}\}_{k=0}^{N-1}\in\mathbb{R}^N$.

Let $\mathbf{U}$ be a unitary matrix. Since unitary matrices do not change the length of a vector, we have
\begin{equation*}
\begin{split}
f(\mathbf{u})&=\frac{1}{2}\sum_{m=0}^{\left\lceil \frac{N}{L} \right\rceil-1} \| \mathbf{U}\left(\mathbf{\tilde{H}}_m- \mathbf{Z}_m\right)\|_2^2 \\
&=\frac{1}{2}\sum_{m=0}^{\left\lceil \frac{N}{L} \right\rceil-1} \| \mathbf{U}\mathbf{\tilde{H}}_m- \mathbf{U}\mathbf{Z}_m\|_2^2.
\end{split}
\end{equation*}
By choosing $\mathbf{U}$ to be the DFT matrix and normalize, we get exactly the loss function in (\ref{eq:LS}).


\subsection{\label{sec:proof_of_interpolation}Proof of Lemma \ref{claim_interpolation}}

We identify the convolution $\mathbf{g}\ast\mathbf{x}$ by the matrix-vector product $\mathbf{G}\mathbf{x}$, where $\mathbf{G}\in\mathbb{R}^{N \times N}$ is a circulant matrix whose first column is given by $\mathbf{\tilde{g}}:=\left\{\mathbf{g}[(-n)\bmod N ]\right\}_{n=0}^{N-1}$. 
For $L=1$, we can then write 
\[
\mathbf{y}=\mathbf{Gx}=\mathbf{F^{*}\Sigma Fx},
\]
where $\mathbf{F}$ is a DFT matrix and $\mathbf{\Sigma}$ is a diagonal
matrix whose entries are the DFT of $\mathbf{\tilde{g}}$. By assumption,
the first $ N/L $ entries of $\mathbf{\Sigma}$ are ones and the rest
are zeros. Hence, we may write 
\begin{equation} 
\mathbf{y}=\mathbf{F}_{p}^{*}\mathbf{F}_{p}\mathbf{x},\tag{E.1}\label{eq:1}
\end{equation}
where $\mathbf{F}_{p}\in\mathbb{C}^{ N/L \times N}$ consists of the
first $ N/L $ rows of $\mathbf{F}$.

Let $\mathbf{G}_{L}\in\mathbb{R}^{ \frac{N}{L} \times N}$ be a matrix consists of the $\left\{ jL\thinspace:\thinspace j=0,\dots, N/L -1\right\} $ rows of $\mathbf{G}$. For $L>1$, we get the downsampled system of equations
\[
\mathbf{y}_{L}=\mathbf{G}_{L}\mathbf{x}=\mathbf{F}_{L}^{*}\mathbf{\Sigma Fx},
\]
where $\mathbf{F}_{L}$ consists of the $\left\{ jL\thinspace:\thinspace j=0,\dots, N/L -1\right\} $
columns of $\mathbf{F}$ (notice the difference between $\mathbf{F}_{L}$
and $\mathbf{F}_{p}$). We aim at showing that expanding and interpolating
$\mathbf{y}_{L}$ as explained in Lemma \ref{claim_interpolation}
results in $\mathbf{y}$. Direct computation shows that the expansion
stage as described in (\ref{eq:expansion}) is equivalent to multiplying
both sides by $\mathbf{F^{*}F}_{L}$: 
\[
\mathbf{\tilde{y}}_{L}=\mathbf{F}^{*}\mathbf{F}_{L}\mathbf{y}_{L}=\mathbf{F}^{*}\left(\mathbf{F}_{L}\mathbf{F}_{L}^{*}\right)\mathbf{\Sigma Fx}.
\]
Let us denote $\mathbf{T:=F}_{L}\mathbf{F}_{L}^{*}$, which is a Toeplitz
matrix with $L$ on the $\frac{jN}{L}$ diagonals for $j=0,\dots, N/L -1$
and zero otherwise. Because of the structure of $\mathbf{\Sigma}$
we can then write 
\[
\mathbf{\tilde{y}}_{L}=\mathbf{F^{*}T}_{p}\mathbf{F}_{p}\mathbf{x},
\]
where $\mathbf{T}_{p}\in\mathbb{R}^{N\times\frac{N}{L}} $ consists of the
first $ N/L $ columns of $\mathbf{T}$. Direct calculation shows that $\mathbf{F}_{p}\mathbf{F^{*}T}_{p}=\mathbf{I}$,
where $\mathbf{I}$ is the identity matrix. Therefore we conclude
that 
\begin{equation}
\left(\mathbf{F}_{p}^{*}\mathbf{F}_{p}\right)\tilde{\mathbf{y}}_{L}=\mathbf{F}_{p}^{*}\mathbf{F}_{p}\mathbf{x}.\tag{E.2}\label{eq:2}
\end{equation}
Comparing (\ref{eq:2}) with (\ref{eq:1}) completes the proof.


\subsection{\label{sub:Proof-of-Lemma_grad_bound}Proof of Lemma \ref{lem:grad_bound}}

Recall that 
\begin{equation*}
\begin{split}\nabla f(\mathbf{z}) & =\sum_{m=0}^{\left\lceil N/L \right\rceil-1}\sum_{\ell=-\left(W-1\right)}^{W-1}\left(\mathbf{z}^{T}\mathbf{H}_{m,\ell}\mathbf{z}-\mathbf{Y}\left[m,\ell\right]\right)\\
 & \cdot \left(\mathbf{H}_{m,\ell}+\mathbf{H}_{m,\ell}^{T}\right)\mathbf{z},
\end{split}
\end{equation*}
where 
\[
\mathbf{H}_{m,\ell}:=\mathbf{P}_{-\ell}\mathbf{D}_{mL}\mathbf{D}_{mL-\ell},
\]
$\mathbf{D}_{mL}$ is a diagonal matrix whose entries are $\left\{ \mathbf{g}\left[mL-n\right]\right\} _{n=0}^{N-1}$
for fixed $m$ and $\mathbf{P}_{\ell}$ is a matrix that shifts (circularly)
the entries of an arbitrary vector by $\ell$ entries. We observe
that for a rectangular window of length $W$ and $\left\Vert \mathbf{z}\right\Vert _{\infty}\leq\frac{1}{\sqrt{N}}$,
we have $\left\Vert \mathbf{z}\right\Vert _{2}\leq 1$
and 
\begin{eqnarray*}
\left\Vert \mathbf{H}_{m,\ell}\mathbf{z}\right\Vert _{2}  \leq  \left\Vert \mathbf{H}_{m,\ell}\right\Vert _{2}\left\Vert \mathbf{z}\right\Vert _{2}
 \leq  1,
\end{eqnarray*}
 so that 
\begin{equation} \tag{F.1}
\left\Vert \nabla f(\mathbf{z})\right\Vert _{2}\leq2\sum_{m=0}^{\left\lceil N/L \right\rceil-1}\sum_{\ell=-\left(W-1\right)}^{W-1}\left|\mathbf{Y}\left[m,\ell\right]-\mathbf{z}^{T}\mathbf{H}_{m,\ell}\mathbf{z}\right|.\label{eq:grad_bound}
\end{equation}

 For convenience, let us denote $d\left(\mathbf{x},\mathbf{z}\right)=\frac{\varepsilon}{\sqrt{N}}$ for some $\varepsilon\leq 1$ and therefore  $\vert \mathbf{z}[n]\vert \geq\frac{1- \varepsilon}{\sqrt{N}}$
for all $n$. Accordingly, for any $(n,k)$,
 \[ \frac{(1-\varepsilon)^2}{N}\leq\vert \mathbf{z}[n]\mathbf{z}[n+k]\vert\leq\frac{1}{N}.\]
 Since $\mathbf{x}[n]$ and $\mathbf{z}[n]\vert$ have the same sign pattern, we have
\begin{equation*}
\begin{split}
\left \vert\mathbf{x}[n]\mathbf{x}[n+k]-\mathbf{z}[n]\mathbf{z}[n+k]\right\vert &\leq \frac{1}{N}\left \vert 1- (1-\varepsilon)^2\right\vert \\
 & \leq \frac{2\varepsilon}{N},
 \end{split} 
\end{equation*}
and  for all $m,\ell\geq 0$, 
\begin{equation}\tag{F.2}
\begin{split}&\left|\mathbf{Y}\left[m,\ell\right]-\mathbf{z}^{T}\mathbf{H}_{m,\ell}\mathbf{z}\right|\\ & \leq \sum_{k=m-(W-1)}^{m-\ell}\left \vert\mathbf{x}[n]\mathbf{x}[n+k]-\mathbf{z}[n]\mathbf{z}[n+k]\right\vert  \\  & \leq\sum_{k=m-(W-1)}^{m-\ell}\frac{2\varepsilon}{N} \leq \frac{2W\varepsilon}{N}  
=\frac{2Wd(\mathbf{x},\mathbf{z})}{\sqrt{N}}.\label{eq:grad_bound2}
\end{split}
\end{equation}
The same bound holds for $\ell<0$. Combining (\ref{eq:grad_bound}) and
(\ref{eq:grad_bound2}) we conclude that 
\begin{eqnarray*}
\left\Vert \nabla f(\mathbf{z})\right\Vert _{2} & \leq & 2\sum_{m=0}^{\left\lceil N/L \right\rceil-1}\sum_{\ell=-\left(W-1\right)}^{W-1}\frac{2Wd(\mathbf{x},\mathbf{z})}{\sqrt{N}}\\
 & = & \frac{8}{L}W^{2}\sqrt{N}d(\mathbf{x},\mathbf{z}).
\end{eqnarray*}

\begin{rem*}
\label{rem:grad_window}In case of a non-vanishing window of length
$W$, one can easily bound the gradient using the same technique,
while taking into account $\max_{n}\left|\mathbf{g}[n]\right|$ in the inequalities.
\end{rem*}


\subsection{\label{sub:Proof-of-Lemma_regularity_con}Proof of Lemma \ref{lem:regualrity_con}}

Recall that (see (\ref{eq:grad})) 

\begin{flalign*}
 &\left\langle \nabla f(\mathbf{z}),\mathbf{z}-\mathbf{x}e^{j\phi(\mathbf{z})}\right\rangle  \\  & =  \sum_{m=0}^{\left\lceil N/L \right\rceil-1}\sum_{\ell=-\left(W-1\right)}^{W-1}\left(\mathbf{z}^{T}\mathbf{H}_{m,\ell}\mathbf{z}-\mathbf{x}^{T}\mathbf{H}_{m,\ell}\mathbf{x}\right) \\ &\cdot 
  \left(\mathbf{z}-\mathbf{x}e^{j\phi(\mathbf{z})}\right)^{T}\left(\mathbf{H}_{m,\ell}+\mathbf{H}_{m,\ell}^{T}\right)\mathbf{z}.\label{eq:explicit_grad}  
\end{flalign*}
Since $\mathbf{x}^T\mathbf{H}_{m,\ell}^T\mathbf{z}=\mathbf{z}^T\mathbf{H}_{m,\ell}\mathbf{x}$ we have for fixed $(m,\ell)$ and  $\phi(\mathbf{z})\in \{0,\pi\}$:
\begin{flalign*}
&\left(\mathbf{z}-\mathbf{x}e^{j\phi(\mathbf{z})}\right)^{T}\left(\mathbf{H}_{m,\ell}+\mathbf{H}_{m,\ell}^{T}\right)\mathbf{z} \\  &=\left(\mathbf{z}-\mathbf{x}e^{j\phi(\mathbf{z})}\right)^{T}\mathbf{H}_{m,\ell}\left(\mathbf{z}-\mathbf{x}e^{j\phi(\mathbf{z})}\right)  +\left(\mathbf{z}^{T}\mathbf{H}_{m,\ell}\mathbf{z}-\mathbf{x}^{T}\mathbf{H}_{m,\ell}\mathbf{x}\right).
\end{flalign*}
Therefore, 
\begin{equation} \tag{G.1}
\begin{split}
\left \langle \nabla f(\mathbf{z}),\mathbf{z}-\mathbf{x}e^{j\phi(\mathbf{z})}\right\rangle & \\ =&  \sum_{m=0}^{N-1}\sum_{\ell=-\left(W-1\right)}^{W-1}\left(\mathbf{z}^{T}\mathbf{H}_{m,\ell}\mathbf{z}-\mathbf{x}^{T}\mathbf{H}_{m,\ell}\mathbf{x}\right)^{2} \\
 +&  \sum_{m=0}^{N-1}\sum_{\ell=-\left(W-1\right)}^{W-1}\left(\mathbf{z}^{T}\mathbf{H}_{m,\ell}\mathbf{z}-\mathbf{x}^{T}\mathbf{H}_{m,\ell}\mathbf{x}\right)   \\  \cdot&
   \left(\mathbf{z}-\mathbf{x}e^{j\phi(\mathbf{z})}\right)^{T}\mathbf{H}_{m,\ell}\left(\mathbf{z}-\mathbf{x}e^{j\phi(\mathbf{z})}\right).\label{eq:grad_identity}
\end{split}
\end{equation}
 Clearly, if $\mathbf{z}=\mathbf{x}e^{j\phi(\mathbf{z})}$ then $\left\langle \nabla f(\mathbf{z}),\mathbf{z}-\mathbf{x}e^{j\phi(\mathbf{z})}\right\rangle =0$.
Otherwise, the first term of (\ref{eq:grad_identity}) is strictly
positive. Hence, in order to achieve a lower bound on (\ref{eq:grad_identity}),
we first derive an upper bound on the second term and then bound the first term from below.

By assumption $d(\mathbf{x},\mathbf{z})\leq \frac{1}{\sqrt{N}}$. Denote $\left|\mathbf{x}[n]e^{j\phi(\mathbf{z})}-\mathbf{z}[n]\right|:=\frac{\varepsilon_{n}}{\sqrt{N}}$
for some $\varepsilon_{n}\leq 1$. We observe that $\sum_{n}(\frac{\varepsilon_{n}}{\sqrt{N}})^{2}=d^{2}\left(\mathbf{x},\mathbf{z}\right)$.
For fixed $\ell\geq0$, we can use the Cauchy-Schwarz inequality
to obtain: 
\[
\begin{split}  \sum_{m=0}^{N-1}\left(\mathbf{z}-\mathbf{x}e^{j\phi(\mathbf{z})}\right)^{T}&\mathbf{H}_{m,\ell}\left(\mathbf{z}-\mathbf{x}e^{j\phi(\mathbf{z})}\right)\\
 = &\sum_{m=0}^{N-1}\sum_{n=m-(W-1)}^{m-\ell}\left(\mathbf{z}\left[n\right]-\mathbf{x}\left[n\right]e^{j\phi(\mathbf{z})}\right)\\ \cdot  & \left(\mathbf{z}\left[n+\ell\right]-\mathbf{x}\left[n+\ell\right]e^{j\phi(\mathbf{z})}\right)\\
 \leq & W\sqrt{\sum_{m=0}^{N-1}\frac{\varepsilon_{m}^{2}}{N}}\sqrt{\sum_{m=0}^{N-1}\frac{\varepsilon_{m+\ell}^{2}}{N}} = Wd^{2}(\mathbf{x},\mathbf{z}).
\end{split}
\]
The same bound holds for $\ell<0$. Combining the last result with
(\ref{eq:grad_bound2}) we get for the second term in (\ref{eq:grad_identity})
that 
\begin{equation}\tag{G.2}
\begin{split} & \sum_{m=0}^{N-1}\sum_{\ell=-\left(W-1\right)}^{W-1}\left(\mathbf{z}^{T}\mathbf{H}_{m,\ell}\mathbf{z}-\mathbf{x}^{T}\mathbf{H}_{m,\ell}\mathbf{x}\right)\cdot\\
 & \left(\mathbf{z}-\mathbf{x}e^{j\phi(\mathbf{z})}\right)^{T}\mathbf{H}_{m,\ell}\left(\mathbf{z}-\mathbf{x}e^{j\phi(\mathbf{z})}\right)\leq\frac{4}{\sqrt{N}}W^{3}d^{3}(\mathbf{x},\mathbf{z}).\label{eq:second_term}
\end{split}
\end{equation}

Next, we aim to bound the first term of (\ref{eq:grad_identity})
from below as follows: 
\begin{equation} \tag{G.3}
\begin{split}  \sum_{m=0}^{N-1}\sum_{\ell=-\left(W-1\right)}^{W-1}&\left(\mathbf{z}^{T}\mathbf{H}_{m,\ell}\mathbf{z}-\mathbf{x}^{T}\mathbf{H}_{m,\ell}\mathbf{x}\right)^{2}\\\geq
  \sum_{m=0}^{N-1}&\left(\mathbf{z}^{T}\mathbf{H}_{m,0}\mathbf{z}-\mathbf{x}^{T}\mathbf{H}_{m,0}\mathbf{x}\right)^{2}\\=
  \sum_{m=0}^{N-1}&\left(\sum_{n=m-(W-1)}^{m}\mathbf{z}^{2}\left[n\right]-\mathbf{x}^{2}\left[n\right]\right)^{2}\\\geq
  \sum_{m=0}^{N-1}&\sum_{n=m-(W-1)}^{m}\left(\mathbf{z}^{2}\left[n\right]-\mathbf{x}^{2}\left[n\right]\right)^{2},\label{eq:lower_bound}
\end{split}
\end{equation}
where the last inequality is true since 
$\mathbf{x}^{2}[n]\geq\mathbf{z}^{2}[n]$ and for any positive (or
negative) sequence $\left\{ a_{i}\right\} $ we have $\left(\sum_{i}a_{i}\right)^{2}\geq\sum_{i}a_{i}^{2}$. Furthermore, since $\vert \mathbf{z}[n]\vert=\frac{1-\varepsilon_n}{\sqrt{N}}$ we have
\begin{equation*}
\begin{split} 
\sum_{m=0}^{N-1}\sum_{n=m-(W-1)}^{m}&\left(\mathbf{z}^{2}\left[n\right]-\mathbf{x}^{2}\left[n\right]\right)^{2} \\ =
\frac{1}{N^2}\sum_{m=0}^{N-1}&\sum_{n=m-(W-1)}^{m}\left(1-(1-\varepsilon_n)^2\right)^2 
\\=  \frac{W}{N^{2}}\sum_{n=0}^{N-1}&\left(2\varepsilon_{n}-\varepsilon_{n}^{2}\right)^{2}.
\end{split}
\end{equation*}
Therefore, since $\varepsilon_{n}\leq1$ for
all $n$ we conclude that  
\begin{equation}\tag{G.4}
\begin{split}  \sum_{m=0}^{N-1}\sum_{\ell=-(W-1)}^{W-1}&\left(\mathbf{z}^{T}\mathbf{H}_{m,\ell}\mathbf{z}-\mathbf{x}^{T}\mathbf{H}_{m,\ell}\mathbf{x}\right)^{2}\\
 &\geq\frac{W}{N^{2}}\sum_{n=0}^{N-1}\varepsilon_{n}^{2}=
  \frac{Wd^{2}\left(\mathbf{x},\mathbf{z}\right)}{N}.\label{eq:bound_below1}
\end{split}
\end{equation}
Plugging (\ref{eq:second_term}) and (\ref{eq:bound_below1}) into
(\ref{eq:grad_identity}) yields

\begin{eqnarray*}
	\left\langle \nabla f(\mathbf{z}),\mathbf{z}-\mathbf{x}\right\rangle   &\geq  & \frac{Wd^{2}\left(\mathbf{x},\mathbf{z}\right)}{N}\left(1-4\sqrt{N}W^{2}d(\mathbf{x},\mathbf{z})\right)\\
	&  \geq & \frac{Wd^{2}\left(\mathbf{x},\mathbf{z}\right)}{2N},
\end{eqnarray*}
where the last inequality holds for $d(\mathbf{x},\mathbf{z})\leq\frac{1}{8\sqrt{N}W^{2}}.$

\begin{rem*}
\label{rem:inner_prod}Observe that the analysis for non-vanishing
windows of length $W$ requires only a small modification. In this
case, one should use the maximal and the minimal values of the window
in the above inequalities. For instance, one would need to take $\mbox{g}_{\mbox{min}}:=\min_{n=0,\dots,W-1}\left|\mathbf{g}[n]\right|$ 
into account in~\eqref{eq:lower_bound}. 
\end{rem*}

\end{document}